\documentclass[11pt]{article}

\PassOptionsToPackage{table}{xcolor}
\usepackage[final]{acl}

\usepackage{times}
\usepackage{latexsym}

\usepackage[T1]{fontenc}

\usepackage[utf8]{inputenc}

\usepackage{microtype}

\usepackage{inconsolata}

\usepackage{graphicx}
\usepackage{amsmath}
\usepackage{amssymb}
\usepackage{booktabs}
\usepackage{multirow}
\graphicspath{{../tex/}}

\newcommand{\newupcite}[1]{\citep{#1}}

\title{SimulS2ST-Omni: Data-Efficient Streaming Speech-to-Speech Translation via Explicit Trajectory Supervision}

\author{
  \textbf{Rongshen He\textsuperscript{1,2}},
  \textbf{Xinyu Liang\textsuperscript{1,2}},
  \textbf{Dekun Chen\textsuperscript{1}},
  \textbf{Jiaqi Li\textsuperscript{1}},
  \textbf{Mingjie Chen\textsuperscript{1,2}},
  \textbf{Zhizheng Wu\textsuperscript{1,2,*}}
\\
\\
  \textsuperscript{1}The Chinese University of Hong Kong, Shenzhen
  \\
  \textsuperscript{2}Amphion Technology Co., Ltd.
  \\
  \texttt{rongshenhe@link.cuhk.edu.cn, wuzhizheng@cuhk.edu.cn}
  \\
  \textsuperscript{*}Corresponding author
}

\begin{document}
\maketitle
\begin{abstract}
Long-form streaming speech-to-speech translation (S2ST) requires incremental, unbounded translation under strict latency constraints. Existing methods typically suffer from sentence-bounded supervision or demand massive paired-S2ST supervision. We introduce a training recipe enabling a speech language model for sentence-level and long-form streaming S2ST using only $\sim$2k hours of paired cross-lingual S2ST data, layered atop auxiliary supervision. Anchored by auxiliary multitask training, our approach remains robust even when the paired-S2ST budget itself is reduced by 90\%. Our core contribution, joint text-code trajectory supervision, schedules target text and acoustic semantic codes as a unified commitment path, eliminating the need for separate, unstable speech-side emission controllers. Furthermore, our two-stream Thinker--Talker factorization significantly outperforms unified-decoder baselines by decoupling linguistic reasoning from dense acoustic prediction to mitigate modality interference. Finally, our system achieves highly competitive quality-latency trade-offs on RealSI and ACL60/60-dev, matching state-of-the-art, closed-source S2ST systems such as LiveInterpret~2.0 on ASR-BLEU. 
\\
{\textbf{Project page:}\,\,\url{https://hasaki321.github.io/SimulS2ST-Omni.demo/}}
\end{abstract}



\section{Introduction}
\label{sec:introduction}

Real-time, cross-lingual communication demands systems capable of translating continuous source speech into target speech with high fidelity and low latency. However, current Speech-to-Speech Translation (S2ST) approaches struggle to meet these demands efficiently. Cascaded systems suffer from inherent latency bottlenecks and compounding errors~\cite{1597243}, while end-to-end Speech Language Models (SLMs) remain largely restricted to offline, sentence-bounded processing~\cite{cheng2025uniss,dong2023polyvoicelanguagemodelsspeech,deng2025simuls2s}. Although continuous streaming S2ST systems have recently emerged, they rely heavily on massive synthetic supervision — often exceeding 40k hours of paired audio, or require complex, expert-annotated reinforcement learning~\cite{labiausse2025highfidelitys2st,labiausse2026simultaneous,cheng2025seedliveinterpret20endtoend}. Thus, achieving robust, long-form streaming S2ST under \emph{limited} paired supervision remains a fundamental, unresolved challenge.

This data bottleneck is exacerbated by two deep structural obstacles. First, long-form streaming is inherently a \emph{latent path problem}. A model must autonomously navigate a read/wait/write policy without explicit step-by-step supervision. While trajectory-based methods address this for text-output streaming~\cite{ouyang2025infinisst,wang2025conversational,yu2025simulpl}, they fail for speech output, which requires the irrevocable, time-sensitive commitment of acoustic semantic codes. Second, the model must learn source-speech understanding, target-text planning, and dense target-code prediction from a very small amount of end-to-end evidence. Recent studies show that unified understanding and generation impose conflicting representation requirements~\cite{shi2025balancing,xu2025modality}. Decoupling these pathways or modules has therefore emerged as a necessary strategy to alleviate this conflict~\cite{wu2024janusdecouplingvisualencoding,chen2024octaviusmitigatingtaskinterference}.

This paper overcomes these limitations through three strategic design choices. First, to maximize the efficiency of paired-S2ST supervision specifically---the scarcest resource in this setting---we construct only $\sim$2,000 hours of high-quality Chinese-English paired {S2ST} data from public Automatic Speech Recognition (ASR) and Speech-to-Text Translation (S2TT) corpora via cross-lingual alignment and monotonicity filtering, layered atop large-scale auxiliary ASR/S2TT/MT/TTS supervision that is comparatively abundant. Second, we introduce a \emph{joint} commitment path — extending trajectory supervision to schedule target text and acoustic semantic codes simultaneously. This eliminates the need for unstable, separate speech-side emission controllers. Third, we establish that backbone architecture is paramount. We demonstrate that a two-stream Thinker--Talker factorization significantly mitigates interference between linguistic planning and dense code prediction, outperforming unified-decoder baselines under paired-S2ST data constraints.

We make the following contributions:
\begin{enumerate}
    \item We deliver a streaming S2ST recipe capable of long-form streaming S2ST using only $\sim$2k hours of filtered paired-S2ST data, or less, by leveraging auxiliary multitask supervision.
    \item We introduce a unified commitment path that schedules text and acoustic codes simultaneously without relying on closed-loop rewards or external policies.
    \item We conduct controlled experiments showing that the two-stream architecture outperforms a unified-decoder baseline for speech translation under low-resource S2ST.
\end{enumerate}

\section{Related Work}
\label{sec:related-work}

Research on Speech-to-Speech Translation (S2ST) has increasingly shifted from offline systems~\newupcite{inaguma-etal-2023-unity,communication2023SeamlessM4T} toward streaming models that minimize latency. Streaming S2ST requires a model to autonomously decide both \emph{what} to generate and \emph{when} to commit based on incomplete speech input. Earlier task-specific streaming architectures typically relied on sentence-bounded cues, such as CTC alignment triggers~\newupcite{zhang2024StreamSpeecha}, transducer-style structures~\newupcite{ma2024overcoming}, or customized monotonic attention and latent-alignment segmentation~\newupcite{communication2023seamlessmultilingualexpressivestreaming,polak2024long}. Although effective for short-form, pre-segmented utterances, these boundary-reliant methods fail to provide a stable training interface for continuous, unsegmented long-form speech. Furthermore, their limited reasoning capacities fall far short of modern large language models (LLMs).

Consequently, recent work has adopted Speech Language Models (SLMs) to inherit these advanced reasoning capabilities, generally clustering into unified-decoder~\newupcite{wang2023viola,zeng2024glm4voice,cheng2025uniss} and decoupled two-stream~\newupcite{dong2023polyvoicelanguagemodelsspeech,xu2025qwen25omni} architectures. While state-of-the-art streaming SLMs such as Hibiki~\newupcite{labiausse2025highfidelitys2st,labiausse2026simultaneous} and LiveInterpret~2.0~\newupcite{cheng2024towards,cheng2025seedliveinterpret20endtoend} successfully achieve unbounded speech-output streaming, they rely heavily on massive volumes of paired source-target audio, expert-annotated alignments, or complex reinforcement learning.

To mitigate the severe scarcity of paired-S2ST data, multiple workarounds have been proposed. Some approaches reduce dependence on paired audio by leveraging multitask or self-supervised representations~\newupcite{lee-etal-2022-textless,zheng2026rosettaspeechzeroshotspeechtospeechtranslation,fang2024can,nachmani2024translatotron}. Others, such as SimulS2S-LLM~\newupcite{deng2025simuls2s} and SimulU~\newupcite{djanibekov2026simulu}, bypass it entirely through test-time wait-$k$ policies or training-free cross-attention decision rules. However, these mitigation strategies remain highly sensitive to hyperparameters and heavily dependent on offline, pre-segmented alignments, preventing them from scaling efficiently to continuous, long-form streams.

To exploit the sequential reasoning of LLMs for unbounded continuous input, a complementary line of research models streaming as a multi-turn dialogue or as a long-form, text-only commitment path~\newupcite{wang2025conversational,guo2025streamuni}. Systems like InfiniSST~\newupcite{ouyang2025infinisst} successfully manage continuous streams by embedding the streaming policy directly into the autoregressive LLM, thereby leveraging its native pretraining intelligence. We bridge the remaining gap between these text-only trajectories and actual speech generation by introducing a \emph{joint text-code commitment path}. By providing token-level joint supervision of text and acoustic codes along time-aligned boundaries, our approach fully preserves the reasoning capability of the LLM. This enables robust, long-form speech-output streaming under a highly constrained paired-data budget, eliminating the need for a separate speech-side emission controller~\newupcite{sudoh2020simultaneous,le2025simulmega}.

\begin{figure*}[!h]
  \centering
  \includegraphics[width=0.98\textwidth]{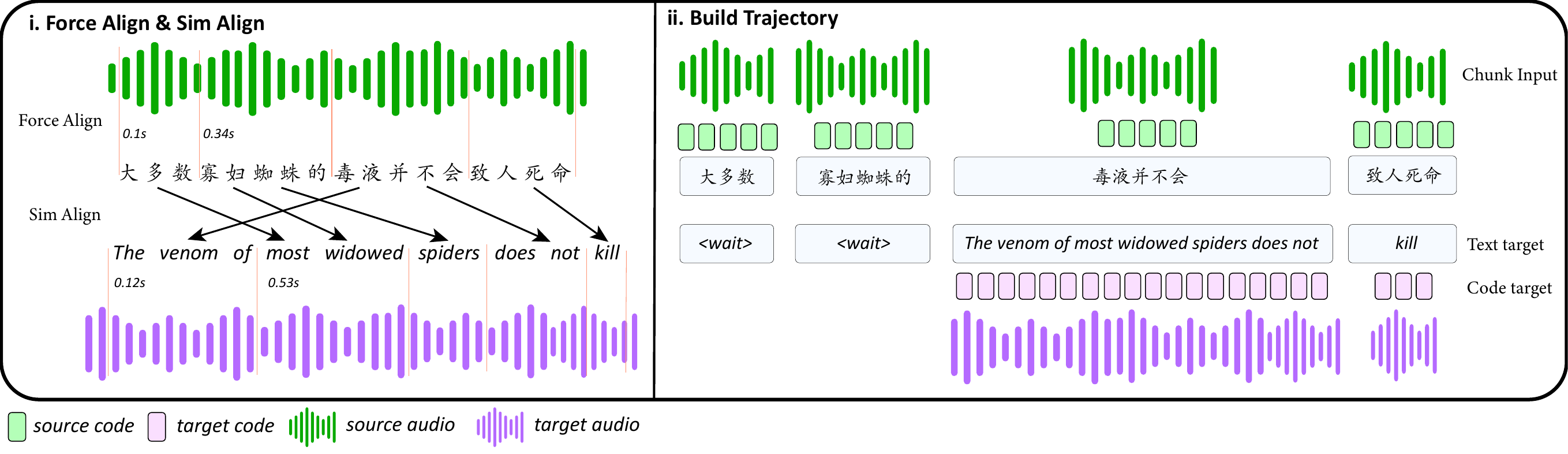}
  \caption{\textbf{Streaming trajectory construction.} \textbf{Step 1:} Word-level and cross-lingual alignments establish the earliest valid source prefix for each target word. \textbf{Step 2:} Target text and segmented target speech codes inherit monotonized boundaries and are grouped into discrete $\texttt{read}$/$\texttt{wait}$/$\texttt{write}$ steps for joint emission.}
  \label{fig:method-s2s-trajectory}
\end{figure*}

\section{Method}
\label{sec:methods}

\subsection{Problem Setup and Formulation}
\label{sec:method-formulation}

A paired-audio S2ST dataset consists of $N$ training samples, denoted as $\{(\mathbf{X}_n, \mathbf{Y}_n^{\mathrm{text}}, \mathbf{Y}_n^{\mathrm{wav}})\}_{n=1}^N$, where $\mathbf{X}_n$ is the source speech, $\mathbf{Y}_n^{\mathrm{text}}$ is the target text, and $\mathbf{Y}_n^{\mathrm{wav}}$ is the target speech. For notational brevity, we omit the sample index $n$ in the following sections. 

Instead of modeling continuous waveforms directly, we represent the target speech using discrete semantic codes. Let $\mathbf{Y}^{\mathrm{code}} = \mathcal{Q}(\mathbf{Y}^{\mathrm{wav}})$ be the semantic token sequence extracted by a fixed speech tokenizer $\mathcal{Q}$. During inference, the waveform $\mathbf{Y}^{\mathrm{wav}}$ is reconstructed from $\mathbf{Y}^{\mathrm{code}}$ using a pre-trained, frozen flow-matching and vocoder backend $\mathcal{D}$ \newupcite{le2023voiceboxtextguidedmultilingualuniversal,chen2026flexivoice,siuzdak2023vocos}.

To train the model for streaming generation, we reformulate the joint prediction of target text and codes as a chunk-wise commitment process. We define a streaming \textbf{trajectory} $\boldsymbol{\tau} = \{(\mathbf{Y}^{\mathrm{text}}_c, \mathbf{Y}^{\mathrm{code}}_c, g_c)\}_{c=1}^{C}$ that partitions the joint target sequences into $C$ chunks. Each chunk $c$ is aligned to a source prefix bounded by the frame index $g_c$, representing the portion of the source speech the model must process before emitting the $c$-th chunk. These indices form a monotonically non-decreasing sequence $g_1 \le g_2 \le \cdots \le g_C$, where the final index $g_C$ corresponds to the total length of the source speech.

At each emission step $c$, generation is conditioned on the chunk context, which encompasses the consumed source speech along with the previously generated target text and codes:
\begin{equation}
\mathcal{C}_c = \left( \mathbf{X}_{1:g_c}, \mathbf{Y}^{\mathrm{text}}_{<c}, \mathbf{Y}^{\mathrm{code}}_{<c} \right).
\end{equation}
The model is then trained to maximize the chunk-factorized log-likelihood of the joint target sequences given the trajectory:
\begin{equation}
\begin{aligned}
&\log p\!\left( \mathbf{Y}^{\mathrm{text}}, \mathbf{Y}^{\mathrm{code}} \mid \mathbf{X}, \boldsymbol{\tau} \right) \\
&\quad= \sum_{c=1}^{C} \log p\!\left( \mathbf{Y}^{\mathrm{text}}_c, \mathbf{Y}^{\mathrm{code}}_c \mid \mathcal{C}_c \right).
\end{aligned}
\label{eq:method-formulation}
\end{equation}

\subsection{Trajectory Construction}
\label{sec:method-trajectory}

Given a paired-S2ST sample $(\mathbf{X}, \mathbf{Y}^{\mathrm{text}}, \mathbf{Y}^{\mathrm{wav}})$, our goal is to derive a chunk-level trajectory $\boldsymbol{\tau}$ that dictates which source prefix must be read before a corresponding target text and code chunk can be emitted. We construct $\boldsymbol{\tau}$ in the following two steps, as illustrated in Figure~\ref{fig:method-s2s-trajectory}.

\paragraph{Step 1: Text and Audio Alignment.}
Given an S2TT sample $(\mathbf{X}, \mathbf{Y}^{\mathrm{text}})$, we first force-align the source and target speech to obtain word-level boundaries, and apply SimAlign~\newupcite{jalili-sabet-etal-2020-simalign} to extract cross-lingual word alignments $\hat{\mathbf{A}}$. If a target word $y_i$ aligns to a source word $x_{a(i)}$, we identify the source-side end frame of $x_{a(i)}$, denoted as $t_{a(i)}$. This frame index represents the earliest source prefix containing sufficient acoustic information to commit $y_i$. Following trajectory-based simultaneous translation practices, we monotonize this boundary to ensure non-decreasing write positions even under local syntactic reordering:
\begin{equation}
\tilde t_i = \max(\tilde t_{i-1}, t_{a(i)}),
\end{equation}
where $\tilde t_0 = 0$. 

\begin{figure*}[t]
  \centering
  \includegraphics[width=0.98\textwidth]{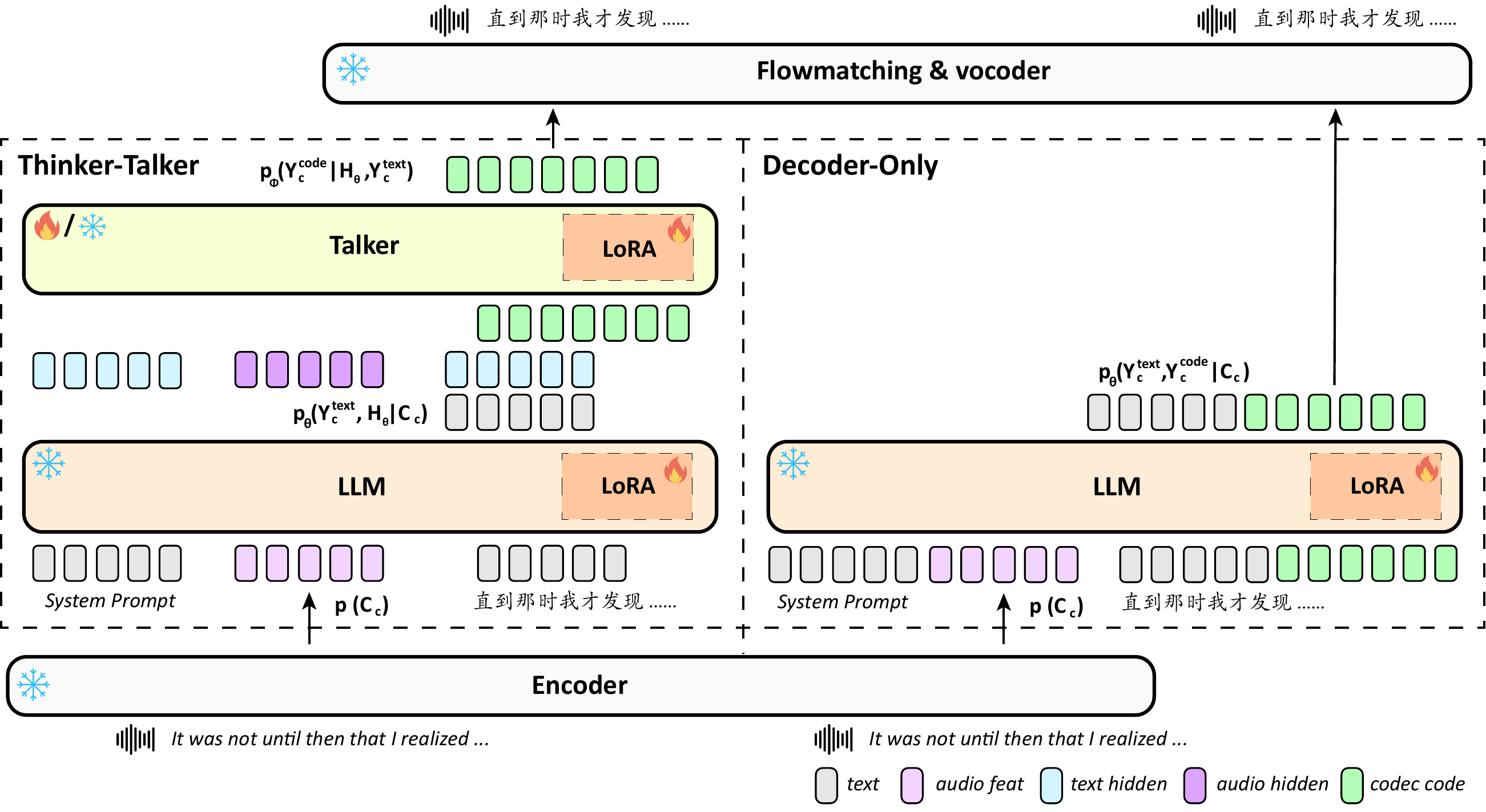}
  \caption{\textbf{Matched backbone comparison.} \textbf{Thinker--Talker} (left) and \textbf{Dec-only} (right) share an identical speech encoder, base LLM backbone (with independent LoRA adapters), semantic-code tokenizer $\mathcal{Q}$, and frozen flow-matching/vocoder backend. 
  }
  \label{fig:method-thinker-talker}
\end{figure*}

\paragraph{Step 2: Code Segmentation and Trajectory Grouping.}
Next, we tokenize the target audio into semantic-code sequences and segment these codes according to the target-word boundaries. Each target word and its corresponding code span inherit the same monotonized source boundary $\tilde t_i$. To form fixed-size streaming steps, we group adjacent target words and their codes whose boundaries fall within the same pre-defined source chunk intervals of \textit{1 second}, yielding the discrete trajectory $\boldsymbol{\tau}$ defined previously. Within this sequence, $g_c$ represents the required source frame boundary before emitting chunk $c$. Chunks without newly committed target content act as \texttt{read}/\texttt{wait} steps; chunks containing target content act as \texttt{write} steps that jointly emit $(\mathbf{Y}^{\mathrm{text}}_c, \mathbf{Y}^{\mathrm{code}}_c)$. Equation~\eqref{eq:method-formulation} represents the resulting chunk-factorized training objective under this teacher trajectory. Notably, offline S2ST is simply the special case where $C=1$ and $g_1=|\mathbf{X}|$, whereas streaming generation corresponds to $C>1$.

\subsection{Backbone Choices}
\label{sec:method-arch}

To isolate the effect of the target-code prediction path under a limited-data regime, we design a controlled comparison between a two-stream architecture and a unified decoder. As shown in Figure~\ref{fig:method-thinker-talker}, both systems are matched in their speech encoder, tokenizer $\mathcal{Q}$, frozen acoustic backend, and training data. They differ exclusively in how they parameterize the chunk-factorized joint distribution $p( \mathbf{Y}^{\mathrm{text}}_c, \mathbf{Y}^{\mathrm{code}}_c \mid \mathcal{C}_c )$ from Equation~\eqref{eq:method-formulation}.

\paragraph{Thinker--Talker.}
The two-stream approach factorizes target-text planning and target-code prediction. The joint probability of the chunk is decomposed as:
\begin{equation}
\begin{aligned}
&p\!\left( \mathbf{Y}^{\mathrm{text}}_c, \mathbf{Y}^{\mathrm{code}}_c \mid \mathcal{C}_c \right) \\
&\quad= p_{\theta}\!\left( \mathbf{Y}^{\mathrm{text}}_c \mid \mathcal{C}_c \right) p_{\phi}\!\left( \mathbf{Y}^{\mathrm{code}}_c \mid \mathcal{C}_c, \mathbf{Y}^{\mathrm{text}}_c, \mathbf{H}_{\theta} \right),
\end{aligned}
\label{eq:method-talker}
\end{equation}
where $\mathbf{H}_{\theta}$ denotes the deterministic sequence of intermediate hidden states produced by the Thinker $\theta$ during text generation. The Thinker $\theta$ and Talker $\phi$ are initialized from Qwen2.5-Omni~\newupcite{xu2025qwen25omni}, with the codec embedding and output head re-initialized for the DualCodec~\cite{li2025dualcodec} vocabulary. The Talker routes code prediction conditioning on the Thinker's hidden states alongside the aligned target-text token embeddings.

\paragraph{Dec-Only Baseline.}
The unified-decoder baseline unifies this process by appending $|\mathcal{V}_{\mathrm{code}}| = 16,384$ code tokens to the Thinker's vocabulary, modeling the joint distribution with a single autoregressive head:
\begin{multline}
p_{\theta}\!\left( \mathbf{Y}^{\mathrm{text}}_c, \mathbf{Y}^{\mathrm{code}}_c \mid \mathcal{C}_c \right), \\
\text{where} \quad \mathcal{V} = \mathcal{V}_{\mathrm{text}} \cup \mathcal{V}_{\mathrm{code}}.
\label{eq:method-deconly}
\end{multline}
Within each chunk $c$, the model autoregressively generates the target text tokens followed immediately by the target code tokens (App.~\ref{app:dec-only}).

\begin{table*}[!t]
  \centering
  \caption{\textbf{Offline S2ST on CVSS-T.} Bold marks the best result among S2ST baselines and our models. ${}^{\dagger}$Scores quoted from \cite{cheng2025uniss}; ${}^{\ddagger}$Re-evaluated under our unified protocol. \texttt{---} denotes undisclosed.}
  \label{tab:cvss}
  \setlength{\tabcolsep}{4pt}
  \resizebox{\textwidth}{!}{%
  \begin{tabular}{llrcccccccc}
    \toprule
     & & & \multicolumn{2}{c}{\textbf{ASR-BLEU}$\uparrow$} & \multicolumn{2}{c}{\textbf{Text-BLEU}$\uparrow$} & \multicolumn{2}{c}{\textbf{A.PCP}$\uparrow$} & \multicolumn{2}{c}{\textbf{SIM-O}$\uparrow$} \\
    \cmidrule(lr){4-5} \cmidrule(lr){6-7} \cmidrule(lr){8-9} \cmidrule(lr){10-11}
    \textbf{Category} & \textbf{Model} & \textbf{\#} & En$\rightarrow$Zh & Zh$\rightarrow$En & En$\rightarrow$Zh & Zh$\rightarrow$En & En$\rightarrow$Zh & Zh$\rightarrow$En & En$\rightarrow$Zh & Zh$\rightarrow$En \\
    \midrule
    \multirow{3}{*}{MLLM}
      & GPT-4o${}^{\dagger}$ & --- & 25.42 & \textbf{31.64} & --- & --- & \textbf{2.66} & 2.58 & --- & --- \\
      & Qwen-Omni2.5${}^{\dagger}$ & 7B & 8.04 & 22.66 & 34.85 & 24.39 & 1.90 & 1.92 & --- & --- \\
      & Step-Audio 2 mini & 7B & \textbf{32.81} & 25.35 & --- & --- & --- & --- & --- & --- \\
    \midrule
    \multirow{4}{*}{S2ST}
      & Seamless-m4t-v2-large${}^{\ddagger}$ & 2.3B & 20.86 & 22.25 & 20.75 & 22.60 & 2.43 & 2.67 & 0.26 & 0.11 \\
      & Seamless-expressive${}^{\ddagger}$ & 1.7B & 23.70 & 15.89 & 25.45 & 16.67 & 2.73 & \textbf{2.76} & 0.42 & 0.34 \\
      & UniSS (P)${}^{\ddagger}$ & 1.5B & 30.09 & 23.77 & 30.77 & 24.63 & 2.73 & 2.75 & 0.40 & 0.43 \\
      & UniSS (Q)${}^{\ddagger}$ & 1.5B & \textbf{32.04} & 24.72 & 32.95 & 25.51 & 2.71 & 2.75 & 0.40 & 0.42 \\
    \midrule
    \multirow{2}{*}{Ours}
      & Ours (Dec-only) & 3B & 27.12 & 23.41 & 31.59 & 24.90 & \textbf{2.96} & 2.75 & \textbf{0.45} & \textbf{0.59} \\
      & Ours (Thinker--Talker) & 3B & 31.12 & \textbf{25.18} & \textbf{33.73} & \textbf{26.41} & 2.70 & 2.64 & 0.35 & 0.45 \\
    \bottomrule
  \end{tabular}%
  }
\end{table*}

\section{Experiments}
\label{sec:experiments}

\subsection{Experimental Protocol}

\begin{table}[!h]
  \centering
  \small 
  \caption{Training data statistics by task type. Token counts are in millions. Audio duration aggregates ASR, S2TT, Stage-1 TTS, and paired-S2ST audio.}
  \label{tab:task_token_stats}
  \begin{tabular}{lrr}
    \toprule
    Task & Tokens (M) & Audio duration (h) \\
    \midrule
    \texttt{TTS}  & 4,531.3 & 89,595.8 \\
    \texttt{MT}  & 4,011.7 & --- \\
    \texttt{ASR}  & 293.1  & 2,811.9 \\
    \texttt{S2TT} & 2597.2  & 24,359.1 \\
    \texttt{S2ST}  & 307.4  & 2,104.8 \\
    \midrule
    \texttt{total} & 11,740.7 & 118,871.6 \\
    \bottomrule
  \end{tabular}
\end{table}

\paragraph{Training Data.}
Our training recipe combines large-scale ASR, S2TT, MT, and TTS supervision with a limited paired-S2ST set (Table~\ref{tab:task_token_stats})---the core low-resource regime studied in this work. Both Thinker--Talker and Dec-only observe the exact same audio data pool. We filter candidate trajectories via length and difficulty quotas using the Normalized Inversion Rate (NIR)~\newupcite{yu2025simulpl} and balance them. Paired-data statistics are in Appendix~\ref{app:data-details}; filtering details are in App.~\ref{app:nir-trajectory} and \ref{app:asr-pair-filtering}.

\paragraph{Matched Three-Stage Training.}
Both backbones share identical data mixtures, optimizers, and token budgets across three stages.\textit{Stage 1 (Warmup)}: Thinker--Talker trains only the Talker on TTS data, while Dec-only uses a dual-branch LoRA~\cite{hu2021loralowrankadaptationlarge} warmup to prevent modality underfitting, drawing inspiration from modality-specialized parameter partitioning~\cite{xu2025modality}.\textit{Stage 2 (Joint Pretraining)}: Both train on an ASR/S2TT/MT/TTS/S2ST mixture at a $0.2{:}1{:}0.5{:}1{:}1.5$ ratio. \textit{Stage 3 (Streaming Finetuning)}: We merge Stage-2 adapters, attach fresh LoRAs, freeze embeddings and prediction heads, and finetune on augmented streaming trajectories consisting of ASR/S2TT/MT/S2ST tasks.
To support all inference latency tiers with a single checkpoint, training trajectories are augmented by sampling a latency multiplier $m \in \{1,\ldots,12\}$, which merges $m$ source chunks into a single read step.
Overall, Dec-only consumes roughly $\boldsymbol{2\times}$ the GPU-hours of Thinker--Talker (Appendix Table~\ref{tab:training_hparams}).

\paragraph{Inference Configuration.}
Offline S2ST decodes full source utterances using greedy sampling, whereas streaming evaluation follows trajectory-defined latency tiers obtained by sweeping the latency multiplier $m$ from a single checkpoint. Streaming decoding uses a rolling-window KV-cache and chunked flow-matching with cross-chunk audio context for speech. Following \newupcite{ouyang2025infinisst}, we use \textit{num\_beams=4} for the main streaming results and greedy decoding for comparison (full hyperparameters in Appendix~\ref{app:inference}).

\paragraph{Evaluation Benchmarks and Metrics.}
We evaluate offline S2ST on CVSS-T~\cite{jia-etal-2022-cvss}. For streaming evaluation, we use RealSI~\cite{cheng2024towards} for sentence-level/long-form S2TT/S2ST, and ACL60/60-dev~\cite{salesky2023evaluating} for long-form En$\rightarrow$Zh streaming S2TT. Crucially, all streaming test sets consist of real-world human recordings; synthesized speech is used strictly to construct the paired-S2ST training data and is never used as an evaluation target. Unless noted otherwise, all benchmarks cover both En$\rightarrow$Zh and Zh$\rightarrow$En. We report BLEU~\cite{post-2018-call} and ASR-BLEU for translation quality, SIM-O and AutoPCP (A.PCP)~\cite{andrews-etal-2022-stopes,cheng2025uniss} for speech quality, and Length-Adaptive Average Lagging (LAAL)~\cite{papi2022over} or StreamLAAL~\cite{papi2024streamatt} for streaming latency. Protocols are in Appendix~\ref{app:offline-eval-details}; metrics and baselines are in Appendix~\ref{app:baseline}. We additionally re-score with a second ASR system (Appendix~\ref{app:asr-sensitivity}).

\begin{figure*}[t]
  \centering
  \includegraphics[width=\linewidth]{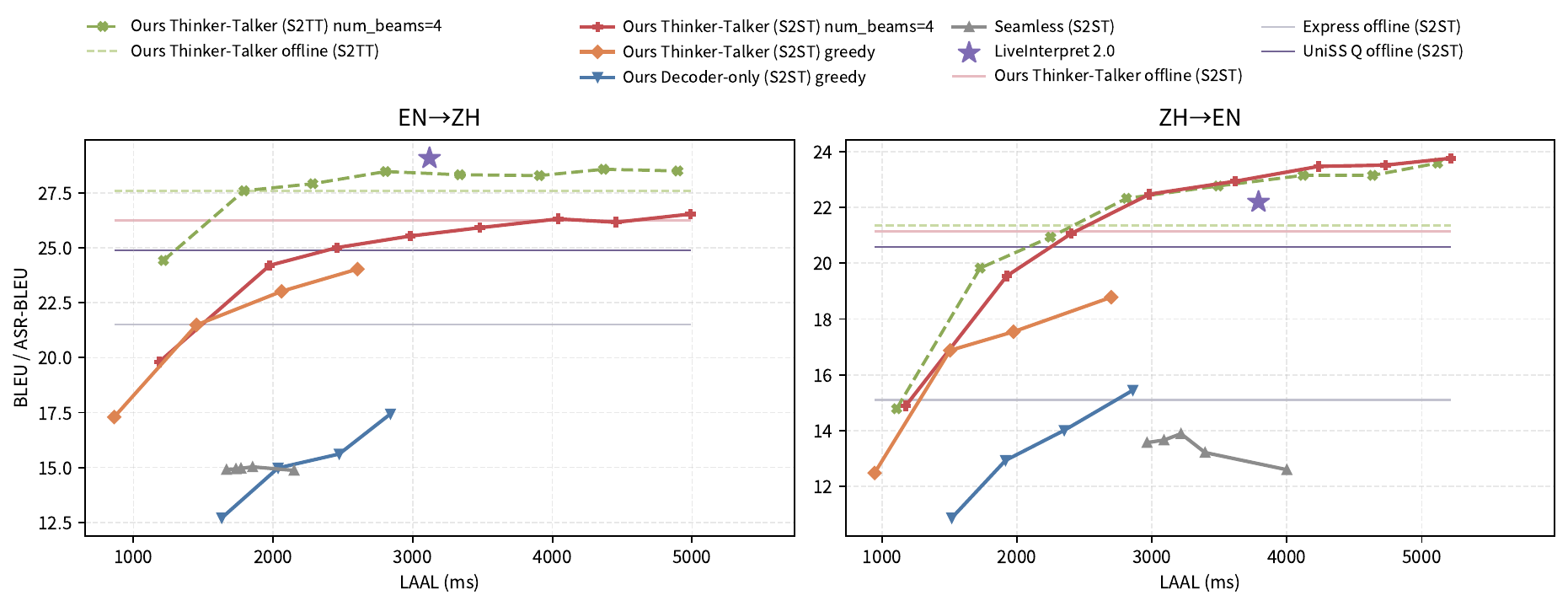}
  \caption{\textbf{RealSI sentence-level content trade-off.} Dashed curves denote S2TT text BLEU, solid curves denote S2ST ASR-BLEU, and horizontal lines mark offline reference lines.}
  \label{fig:realsi_sentence_tradeoff}
\end{figure*}

\begin{figure}[t]
  \centering
  \includegraphics[width=\linewidth]{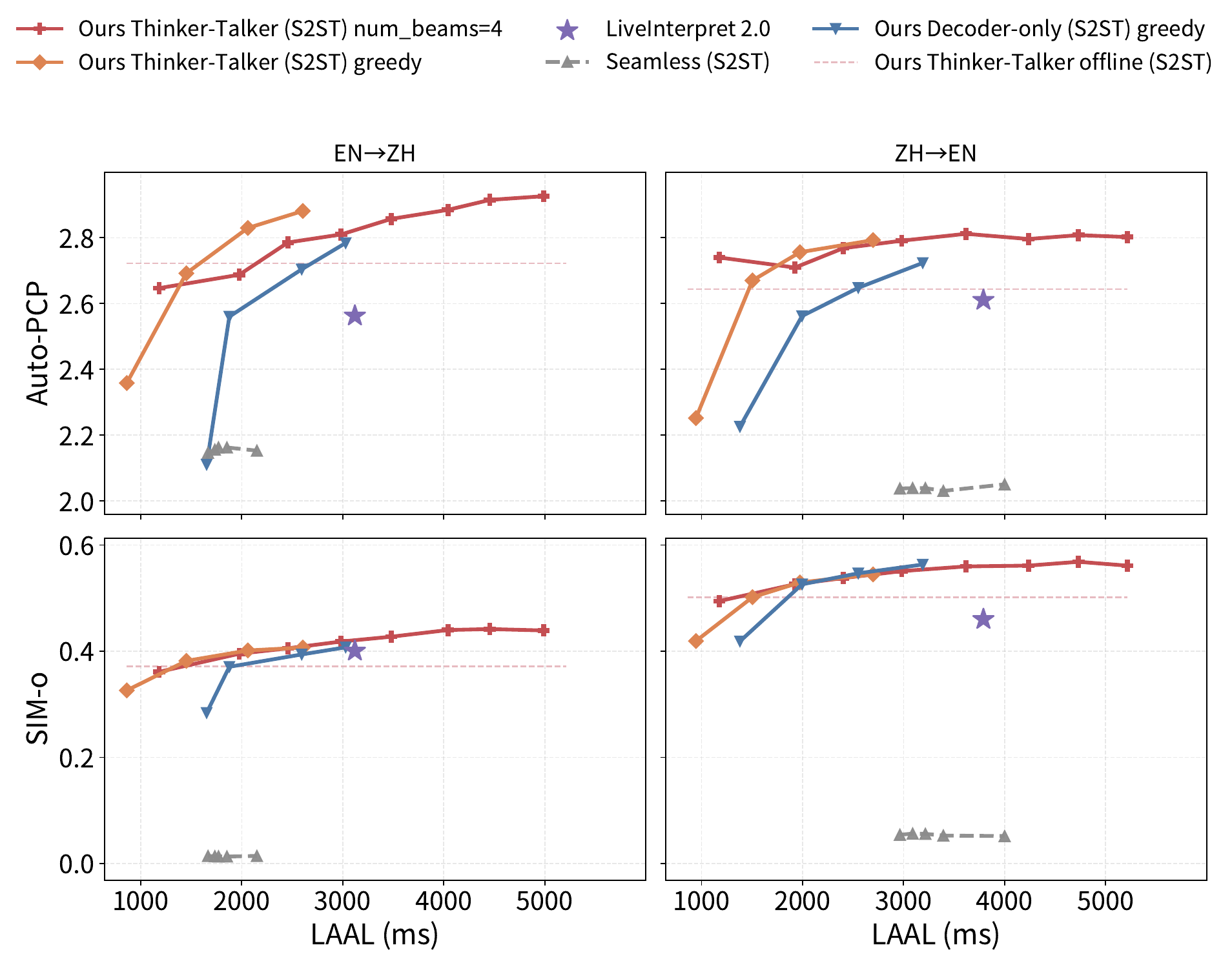}
  \caption{\textbf{RealSI sentence-level acoustic quality trade-off:} A.PCP and SIM-O against LAAL for En$\rightarrow$Zh and Zh$\rightarrow$En.}
  \label{fig:realsi_sentence_quality_tradeoff}
\end{figure}

\subsection{Offline S2ST Evaluation}

We first evaluate the model's offline S2ST performance on the CVSS-T dataset. This reflects the model's basic speech translation and generation abilities aside from the added complexities of streaming, allowing for a clean comparison between the Thinker--Talker and Dec-only architectures. We evaluate them against state-of-the-art public baselines, including GPT-4o~\cite{openai2024gpt4-o}, Qwen-Omni2.5~\cite{xu2025qwen25omni}, Seamless-M4T~\cite{communication2023SeamlessM4T}, and UniSS~\cite{cheng2025uniss} (Table~\ref{tab:cvss}); see also App.~\ref{app:offline-mt-asr}.

As shown in Table~\ref{tab:cvss}, the two-stream Thinker--Talker architecture substantially improves translation quality over the matched Dec-only baseline, successfully matching the performance of the previous state-of-the-art, UniSS(Q). In both directions, the Talker yields higher scores in both Text-BLEU and ASR-BLEU. This confirms that forcing a unified decoder to predict dense semantic codes introduces modality interference that harms intermediate text planning. Furthermore, on En$\rightarrow$Zh, the Talker's +4.0 ASR-BLEU gain notably outpaces its +2.14 Text-BLEU gain. This outsized content-realization improvement indicates that the two-stream design provides a dual benefit: it protects text reasoning and establishes a much more faithful path for target-speech generation. However, the Dec-only baseline maintains a clear edge in non-text acoustic quality, scoring higher on A.PCP and SIM-O and even outperforming UniSS. This highlights a fundamental intelligence-versus-quality trade-off: the Dec-only model leverages its entire 3B backbone for code generation, which strengthens its ability to capture fine-grained audio features but severely harms linguistic intelligence. In contrast, the two-stream system restricts code generation to a lightweight 0.4B Talker, sacrificing some acoustic richness to protect overall translation fidelity prior to trajectory finetuning.

\subsection{Sentence-Level Simultaneous Translation}

Sentence-level simultaneous translation is a bounded setting for evaluating simultaneous S2TT or S2ST. In the resulting frontier plots, the horizontal axis represents the LAAL delay, while the vertical axis denotes translation quality, measured by text BLEU (for S2TT) or ASR-BLEU (for S2ST). A better-performing system yields curves positioned closer to the top-left region, achieving both lower latency and higher translation quality. We evaluate our system on RealSI~\cite{cheng2024towards} against SeamlessStreaming~\cite{communication2023seamlessmultilingualexpressivestreaming}, LiveInterpret~2.0~\cite{cheng2025seedliveinterpret20endtoend}\footnote{Quality scores use audio obtained by using its \href{https://www.volcengine.com/docs/6561/1756902?lang=en}{official Volcengine API script} (concatenated, non-chunked audio), scored with our unified pipeline. Latency uses the LAAL reported in its paper rather than an API-measured value, to avoid service-side queuing delay; refer to Appendix.~\ref{app:ca-eval} for LAAL\_CA analysis.} , and offline models.

Figure~\ref{fig:realsi_sentence_tradeoff} illustrates the sentence-level content trade-off. The two-stream Talker vastly outperforms both Seamless streaming and the matched Dec-only baseline across all latency tiers in both directions. Crucially, this massive ASR-BLEU lead confirms that trajectory finetuning alone cannot resolve a unified decoder's internal text/code conflict, cementing the structural necessity of two-stream factorization for continuous streaming. This persists under CA/RTF analysis, where Thinker--Talker remains stronger and more efficient than Dec-only (App.~\ref{app:ca-eval}). Furthermore, under beam search, the Talker scales so effectively that it eventually surpasses its own offline S2ST reference at higher latency tiers, and from tier $m3$ onward its streaming operating points already exceed UniSS-Q~\cite{cheng2025uniss}, a strong open-source \textit{offline} S2ST system, in both directions---despite UniSS-Q having access to the full utterance while our system decodes under a streaming constraint.

Figure~\ref{fig:realsi_sentence_quality_tradeoff} demonstrates that these massive translation gains do not compromise acoustic fidelity. Unlike in the offline setting, the streaming Talker consistently outperforms Dec-only on A.PCP. More importantly, as the latency multiplier $m$ increases, the Talker's A.PCP scores actually exceed our offline S2ST baseline. This proves that chunked text-code generation via trajectory supervision actively improves local rhythm, rather than merely preserving it. 

Finally, our system achieves highly competitive performance against state-of-the-art, closed-source S2ST systems. Across both directions, the Talker eclipses LiveInterpret~2.0 in speech quality at comparable latencies (Figure~\ref{fig:realsi_sentence_quality_tradeoff}). In translation fidelity (Table~\ref{tab:realsi_streaming_s2st_asr_bleu}), the Talker successfully matches or exceeds LiveInterpret~2.0 from tier $m4$ onward on Zh$\rightarrow$En, and shrinks the gap to a highly competitive margin on En$\rightarrow$Zh. To complement these automatic metrics with human judgments of naturalness, fidelity, and translation adequacy, we further report a human listening study and an LLM-as-judge (VIP) evaluation against LiveInterpret~2.0 and Seamless-Streaming in Appendix~\ref{app:human-eval}, which corroborate the same latency-matched ranking.

\subsection{Long-form Simultaneous Translation}

Long-form streaming evaluates a model's ability to maintain discourse context and prevent low-latency error accumulation over unbounded, continuous source streams. We anchor this evaluation with two complementary comparisons: long-form streaming S2ST against LiveInterpret~2.0 on RealSI~\cite{cheng2024towards}, and streaming S2TT against published academic systems---such as CMU25~\cite{ouyang2025cmu}, NAIST-25~\cite{tan-etal-2025-naist}, and the IWSLT baseline VAD---on ACL60/60-dev~\cite{salesky2023evaluating}\footnote{CMU operating points are taken directly from the CMU paper; NAIST-25 and IWSLT VAD curves are digitized from the NAIST paper to serve as trend references.}. For S2TT, we report BLEU against StreamLAAL~\cite{papi2024streamatt}. We report ASR-BLEU for S2ST translation fidelity. (RealSI long-form S2TT trade-offs are reported in Appendix~\ref{app:realsi_doc_s2tt}.)

\begin{table}[htbp]
  \centering
  \caption{\textbf{RealSI streaming S2ST ASR-BLEU.} Comparison against reported LiveInterpret~2.0 operating points. \textbf{Bold} marks the highest score per row. (The table is truncated to m2--m6 due to the page width limitation.)}
  \label{tab:realsi_streaming_s2st_asr_bleu}
  \scriptsize
  \setlength{\tabcolsep}{3pt}
  \resizebox{\columnwidth}{!}{%
  \begin{tabular}{llccc>{\columncolor{black!10}}c c >{\columncolor{black!10}}c}
    \toprule
    & & \multicolumn{5}{c}{\textbf{Ours}} & \multicolumn{1}{c}{\textbf{LiveInterpret}} \\
    \cmidrule(lr){3-7} \cmidrule(lr){8-8}
    \textbf{Type} & \textbf{Direction} & \textbf{m2} & \textbf{m3} & \textbf{m4} & \textbf{m5} & \textbf{m6} & \textbf{ASR-BLEU} \\
    \midrule
    \multirow{2}{*}{Sent}
      & En$\rightarrow$Zh & 24.21 & 25.01 & 25.54 & 25.92 & 26.32 & \textbf{29.09} \\
      & Zh$\rightarrow$En & 19.55 & 21.07 & 22.47 & 22.94 & \textbf{23.47} & 22.19 \\
    \midrule
    \multirow{2}{*}{Doc}
      & En$\rightarrow$Zh & 26.58 & 28.42 & 29.69 & \textbf{30.04} & 29.73 & 29.33 \\
      & Zh$\rightarrow$En & 22.40 & 22.62 & 24.37 & 26.53 & 25.60 & \textbf{26.80} \\
    \bottomrule
  \end{tabular}%
  }
\end{table}

As shown in Table~\ref{tab:realsi_streaming_s2st_asr_bleu} (bottom rows), the trajectory-finetuned Talker successfully scales to continuous S2ST without the catastrophic performance degradation typically associated with unbounded generation. On the RealSI long-form evaluation, our system remains highly competitive with LiveInterpret~2.0, a heavily optimized closed-source benchmark. Specifically, our higher latency tiers ($m4$--$m6$) successfully exceed the closed-source baseline on En$\rightarrow$Zh and maintain a tight, approaching margin on Zh$\rightarrow$En. This confirms that the joint text-code commitment path effectively manages long-form target speech generation.

\begin{figure}[t]
  \centering
  \includegraphics[width=\columnwidth]{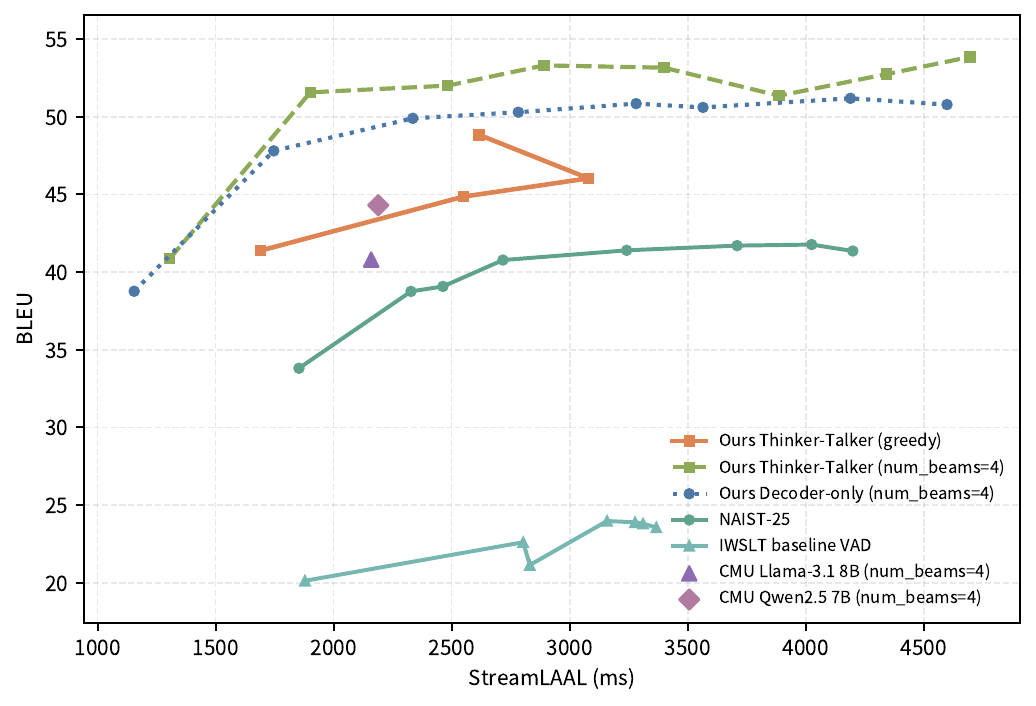}
  \caption{\textbf{ACL60/60-dev long-form En$\rightarrow$Zh streaming S2TT trade-off.} BLEU against StreamLAAL.}
  \label{fig:acl6060_doc_tradeoff}
\end{figure}

Figure~\ref{fig:acl6060_doc_tradeoff} illustrates long-form En$\rightarrow$Zh streaming S2TT on ACL60/60-dev. Consistent with our sentence-level findings, the two-stream Talker strictly dominates the matched Dec-only baseline across all latency tiers. More importantly, the Talker substantially outperforms recent academic systems. Our greedy Talker configuration alone is already competitive with CMU's heavier 7B models operating under beam search at similar latencies ($\sim$2.4\,s StreamLAAL). Meanwhile, our standard beam-search configuration reaches 52.30 BLEU at matched tier m3, decisively outperforming all external baselines in the same latency region. This provides robust, cross-benchmark evidence that the trajectory-and-Talker recipe yields a state-of-the-art long-form quality--latency frontier.

\subsection{Ablation Studies}
\label{sec:experiments-ablations}

\begin{table}[htbp]
  \centering
  \caption{\textbf{Data quality and trajectory sampling strategy ablation on long-form RealSI S2TT.} Stream-side BLEU by latency tier.}
  \label{tab:realsi_doc_ablation}
  \small
  \setlength{\tabcolsep}{4pt}
  \resizebox{\columnwidth}{!}{%
  \begin{tabular}{llrrrr}
    \toprule
    \textbf{Direction} & \textbf{Variant} & \textbf{m1} & \textbf{m2} & \textbf{m3} & \textbf{m4} \\
    \midrule
    \multirow{4}{*}{En$\rightarrow$Zh}
      & \textsc{Random} & 4.59 & 24.86 & 26.22 & 26.54 \\
      & \textsc{NIR} & 21.14 & 26.44 & 28.00 & \textbf{28.53} \\
      & \textsc{NIR+Linear} & \textbf{22.26} & \textbf{26.94} & \textbf{28.19} & 28.33 \\
      & \textsc{NIR+Gaussian} & 22.08 & 26.91 & 27.60 & 27.90 \\
    \midrule
    \multirow{4}{*}{Zh$\rightarrow$En}
      & \textsc{Random} & 3.56 & 14.55 & 17.32 & 17.08 \\
      & \textsc{NIR} & 11.98 & 18.42 & 19.89 & \textbf{21.73} \\
      & \textsc{NIR+Linear} & \textbf{12.83}  & \textbf{18.83} & \textbf{20.40} & 21.63 \\
      & \textsc{NIR+Gaussian} & 11.95 & 17.78 & 19.87 & 21.46 \\
    \bottomrule
  \end{tabular}%
  }
\end{table}

\paragraph{Data quality and sampling strategy.}
Table~\ref{tab:realsi_doc_ablation} isolates the effects of trajectory data filtering and latency multiplier sampling on streaming translation quality, using a controlled subset of 60k trajectories. For data selection, \textsc{Random} draws trajectories arbitrarily, whereas \textsc{NIR} applies our monotonicity-aware filter. Both use a uniform distribution for latency multipliers during training. To test the effect of multiplier sampling, we also evaluate \textsc{NIR+Linear} and \textsc{NIR+Gaussian}, which apply shaped distributions to bias training toward specific latency tiers (see Appendix~\ref{app:training-details}).

Without monotonicity filtering, low-latency ($m1$) performance collapses entirely, plummeting to 4.59 and 3.56 BLEU in the two directions, respectively. Introducing \textsc{NIR}-based filtering immediately recovers this gap, proving that high-quality, difficulty-controlled trajectory supervision is the primary driver of low-latency robustness. Furthermore, varying the multiplier sampling schedules yields negligible differences compared with the uniform baseline. This confirms that curating a stable, filtered data pool is far more critical than the specific multiplier sampling schedule.

\begin{table}[htbp]
  \centering
  \caption{RealSI sentence-level paired-S2ST budget ablation by latency tier. Values are ASR-BLEU. All results use greedy decoding.}
  \label{tab:realsi_s2s_budget_ablation}
  \small
  \setlength{\tabcolsep}{4pt}
  \resizebox{\columnwidth}{!}{%
  \begin{tabular}{lllrrrr}
    \toprule
    \textbf{Direction} & \textbf{Model} & \textbf{Paired-S2ST budget} & \textbf{m1} & \textbf{m2} & \textbf{m3} & \textbf{m4} \\
    \midrule
    \multirow{5}{*}{En$\rightarrow$Zh}
      & Talker & Full & \textbf{17.30} & \textbf{21.49} & \textbf{23.02} & \textbf{24.04} \\
      & Talker & 10\% & 15.50 & 20.00 & 21.49 & 23.47 \\
      & Talker & 10\% w/o aux & 9.47 & 18.46 & 21.12 & 21.90 \\
      \cmidrule(lr){2-7}
      & Dec-only & Full & 12.70 & \textbf{14.98} & \textbf{15.61} & \textbf{17.44} \\
      & Dec-only & 10\% & \textbf{13.02} & 14.07 & 15.38 & 16.52 \\
      & Dec-only & 10\% w/o aux & 12.96 & 13.97 & 14.22 & 15.93 \\
    \midrule
    \multirow{5}{*}{Zh$\rightarrow$En}
      & Talker & Full & \textbf{12.49} & \textbf{16.88} & \textbf{17.55} & \textbf{18.78} \\
      & Talker & 10\% & 9.76 & 15.57 & 17.53  & 18.59 \\
      & Talker & 10\% w/o aux & 8.45 & 14.62 & 16.91 & 18.24 \\
      \cmidrule(lr){2-7}
      & Dec-only & Full & \textbf{10.88} & \textbf{12.94} & 14.01 & \textbf{15.45} \\
      & Dec-only & 10\% & 10.29 & 12.82 & \textbf{14.16} & 15.12 \\
      & Dec-only & 10\% w/o aux & 10.24 & 11.68 & 13.83 & 15.12 \\
    \bottomrule
  \end{tabular}%
  }
\end{table}

\paragraph{Dataset size and augmented tasks}
Table~\ref{tab:realsi_s2s_budget_ablation} isolates the effect of reducing the paired-S2ST supervision to just 10\%, with and without the fixed auxiliary mixture (simul ASR/S2TT/MT). All models are trained under the \textsc{NIR+Linear} trajectory sampling setting.
The Talker exhibits remarkable data efficiency: when auxiliary data is retained, the 10\% model stays highly competitive with the full-data baseline. However, removing this auxiliary data (\texttt{w/o aux}) triggers a severe degradation, dropping by nearly 8.0 ASR-BLEU at $m1$ on En$\rightarrow$Zh. This demonstrates that the auxiliary mixture successfully anchors the model, allowing it to scale up and maintain streaming S2ST using only a fraction of paired speech. Conversely, the Dec-only baseline remains strictly bounded below the Talker's performance ceiling across all subsets. Notably, removing auxiliary data from the 10\% Dec-only model yields negligible differences. This indicates that the unified decoder is fundamentally bottlenecked by its internal text/code conflict and cannot effectively leverage auxiliary multitask supervision to improve streaming performance.

\section{Conclusion}
\label{sec:conclusion}

This work demonstrates that robust, long-form streaming speech-to-speech translation (S2ST) is achievable even under realistic low-resource conditions where paired S2ST audio is scarce. Through a controlled evaluation, we show that the two-stream Thinker--Talker factorization utilizes limited paired data significantly more effectively than unified-decoder baselines, yielding superior offline and simultaneous translation quality. By combining this architecture with a joint text-code trajectory supervision mechanism, our system successfully scales to long-form streaming with controllable latency tiers. Crucially, our ablations confirm that state-of-the-art low-latency performance requires the synergy of both reliable trajectory filtering and the structural decoupling of linguistic and acoustic generation. Future work will extend this framework to duplex architectures for real-time interactive conversation and scale paired data across scenarios.

\section*{Limitations}
Our experiments are limited to Chinese--English because suitable multilingual streaming S2ST data are still scarce. Extending the same trajectory-based training pipeline to broader multilingual settings is an important direction for future work.

In particular, the paired streaming S2ST data used in this work are constructed through synthesis, alignment, and filtering, rather than collected from naturally recorded simultaneous interpretation. As a result, they may not fully capture all phenomena in real-world interpreting scenarios. Collecting real simultaneous-interpretation speech pairs for training and evaluation is an important direction for future work.

In addition, this work mainly studies the language-model and token-prediction backbone, while keeping the speech encoder and speech-generation backend fixed. The current chunk-wise Flow Matching decoder is not yet fully streaming-native, which may limit inter-chunk coherence and inference efficiency. Extending the acoustic backend like streaming Flow Matching or Multi token predictor to explicitly model streaming history is left to future work.

\section*{Ethical Considerations}
S2ST systems that synthesize target speech raise voice-identity and consent concerns. Although this work does not introduce a new voice-cloning model, generated target speech could still be misused for impersonation if deployed without appropriate safeguards. Any release of models, data, or demos should therefore document synthesis sources, avoid unauthorized voice cloning, and make generated audio identifiable where appropriate.

\section*{Acknowledgments}
This work is partially supported by NSFC Grant 62376237; the 2023 Shenzhen Stability Science Program; and the Program for Guangdong Introducing Innovative and Entrepreneurial Teams (2023ZT10X044).


\bibliography{./bib/camera_ready_clean}

\newpage

\appendix
\section{Data Sources, Pair Construction, and Filtering}
\label{app:data-details}

\begin{table}[htbp]
  \centering
  \caption{Training data sources by task type. Sample counts are in thousands (k). Audio duration is aggregated over speech-containing splits. \texttt{short\_audio\_supplement}, \texttt{emilia\_paired\_*}, and \textit{S2ST} rows are derived internal subsets or filtered paired pools built from the cited source corpora.}
  \label{tab:appendix_data_sources}
  \small
  \setlength{\tabcolsep}{4pt}
  \begin{tabular}{lrr}
    \toprule
    \textbf{Dataset} & \textbf{Samples (k)} & \textbf{Hours} \\
    \midrule
    \multicolumn{3}{l}{\textit{ASR}} \\
    \texttt{common\_voice\_en} & 1143.3 & 1{,}808.2 \\
    \texttt{common\_voice\_zh} & 29.6 & 42.6 \\
    \texttt{librispeech} & 281.2 & 961.1 \\
    \texttt{aishell\_2} & 1{,}009 & 1000.0 \\
    \midrule
    \multicolumn{3}{l}{\textit{S2TT}} \\
    \texttt{gigast} & 7{,}665.2 & 9{,}807.1 \\
    \texttt{mustc\_en-zh} & 358.9 & 596.1 \\
    \texttt{covost\_en-zh} & 288.2 & 428.0 \\
    \texttt{covost\_zh-en} & 7.1 & 11.0 \\
    \texttt{wenetspeech} & 14{,}203.7 & 9{,}732.0 \\
    \texttt{emilia\_en-zh} & 598.4 & 1{,}540.1 \\
    \texttt{emilia\_zh-en} & 875.9 & 2{,}244.8 \\
    \midrule
    \multicolumn{3}{l}{\textit{MT}} \\
    \texttt{ai\_challenger} & 10{,}000.0 & --- \\
    \texttt{WMT} & 23{,}206.6 & --- \\
    \midrule
    \multicolumn{3}{l}{\textit{TTS}} \\
    \texttt{emilia\_en} & 15{,}533.1 & 40{,}714.4 \\
    \texttt{emilia\_zh} & 16{,}954.1 & 42{,}540.5 \\
    \texttt{short\_audio\_supplement} & 2{,}036.0 & 1{,}083.2 \\
    \midrule
    \multicolumn{3}{l}{\textit{TTS-Paired}} \\
    \texttt{emilia\_paired\_en} & 999.0 & 2{,}704.3 \\
    \texttt{emilia\_paired\_zh} & 999.0 & 2{,}553.4 \\
    \midrule
    \multicolumn{3}{l}{\textit{S2ST}} \\
    \texttt{s2st\_en-zh} & 685.9 & 1{,}193.3 \\
    \texttt{s2st\_zh-en} & 484.4 & 911.4 \\
    \bottomrule
  \end{tabular}
\end{table}

\subsection{Data Source Statistics}
Our training data combine large publicly available supervision for speech understanding and text translation with a substantially smaller pool of paired-S2ST examples. The full multitask mixture draws from LibriSpeech~\cite{panayotov2015librispeech}, Common Voice~\cite{ardila-etal-2020-common}, MuST-C~\cite{di-gangi-etal-2019-must}, CoVoST~2~\cite{wang2021covost}, GigaST~\cite{ye2023gigast10000hourpseudospeech}, WenetSpeech~\cite{zhang2022wenetspeech10000hoursmultidomain}, Emilia~\cite{he2024emiliaextensivemultilingualdiverse}, and large-scale text-parallel corpora such as AI Challenger~\cite{aichallenger2017mt} and WMT~\cite{bojar-etal-2017-findings}, together with several short-speech supplements derived from public ASR datasets. Their task roles are reported in the main text; this appendix focuses on paired-S2ST construction and the trajectory-specific filters that matter for the controlled comparisons.

Table~\ref{tab:appendix_data_sources} summarizes the multitask training mixture and the filtered paired-S2ST pools used for trajectory construction. Text-only MT corpora do not contribute audio hours and are therefore marked as not applicable.

\subsection{Paired-S2ST Candidate Construction}

Before entering the candidate pool, each source speech corpus is first cleaned at the transcript level: we cross-check two independent ASR engines (Qwen3-ASR-1.7B and Whisper-large-v3) against the corpus's reference transcript under strict word-error-rate and length thresholds, retaining roughly 80--95\% of utterances depending on corpus quality. This keeps the source transcript---and hence the eventual translation target---accurate before any further construction.

The paired-S2ST pool is built to provide $(\mathbf{X}, \mathbf{Y}^{\mathrm{text}}, \mathbf{Y}^{\mathrm{wav}})$ supervision for the offline S2ST stage. In addition to directly available S2TT-style parallel corpora, we expand the candidate pool with publicly available speech corpora for which target text can be generated reliably. Concretely, we sample additional Chinese and English speech from publicly available speech resources. A strong text translation model, Qwen3.6 35BA3B~\cite{yang2025qwen3technicalreport}, produces target-language text when parallel targets are not already available, yielding a pool of roughly 20,000 hours where every candidate has an accurate source transcript and target text. Each candidate then contains source speech $\mathbf{X}$ and target text $\mathbf{Y}^{\mathrm{text}}$, after which target speech $\mathbf{Y}^{\mathrm{wav}}$ is synthesized.

Target speech is synthesized with OmniVoice~\cite{zhu2026omnivoice} under the same cross-lingual voice-cloning setting used in our evaluation pipeline. This keeps the paired-S2ST supervision aligned with the downstream S2ST objective: the target side must be correct in content while remaining usable for semantic-code learning and speech reconstruction.

\subsection{Monotonicity-Aware Trajectory Selection}
\label{app:nir-trajectory}

For streaming training, each reliable full-utterance pair must be converted into incremental read--write supervision. To reduce unstable long-distance reorderings in the trajectory pool, we apply monotonicity-aware selection based on the normalized inversion rate (NIR)~\cite{yu2025simulpl}. Let the source positions aligned to target words form the sequence $\hat{\mathbf{A}}=[\hat{a}_1,\hat{a}_2,\ldots]$. If $I_{\hat{\mathbf{A}}}$ is the inversion number of this sequence, NIR is defined as
\begin{equation}
\mathrm{NIR}=\frac{2I_{\hat{\mathbf{A}}}}{|\hat{\mathbf{A}}|(|\hat{\mathbf{A}}|-1)}\times 100\%.
\end{equation}
Lower NIR indicates weaker reordering and therefore more stable read--write supervision. In the main trajectory pool, we bucket examples by NIR and sentence length. The main sampling policy uses difficulty weights $\{\text{high}:0.1,\text{mid\_high}:0.3,\text{mid\_low}:0.4,\text{low}:0.2\}$ and length weights $\{\text{short}:0.1,\text{medium}:0.5,\text{long}:0.4\}$.

\begin{figure}[htbp]
  \centering
  \includegraphics[width=\columnwidth]{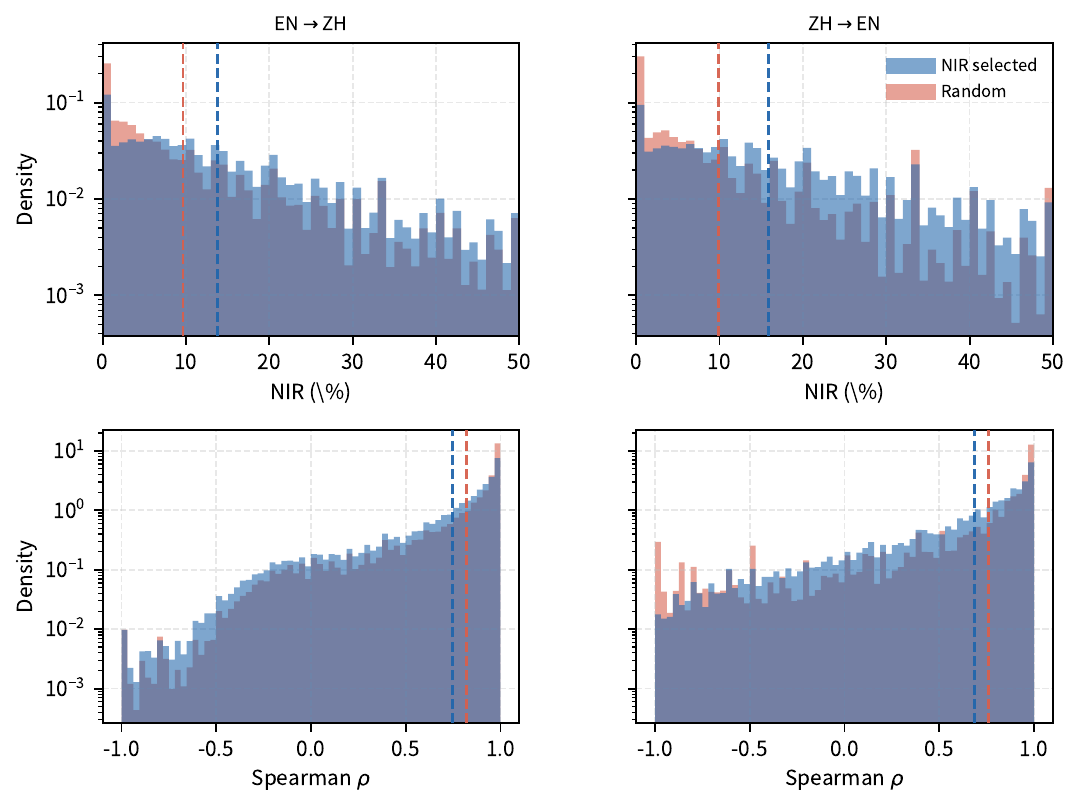}
  \caption{Pooled NIR and Spearman distributions for NIR-stratified vs.\ random trajectory candidates (log-scaled density). Top row: NIR (\%); bottom row: Spearman $\rho$. Columns: En$\rightarrow$Zh (left) and Zh$\rightarrow$En (right). NIR histograms use $\mathrm{NIR}\le 50\%$; dashed lines are means.}
  \label{fig:appendix_nir_random_hist}
\end{figure}

Figure~\ref{fig:appendix_nir_random_hist} and Table~\ref{tab:appendix_nir_random_stats} compare this NIR-stratified trajectory pool against a random-selection baseline drawn from overlapping SimAlign-processed candidates. Metrics are computed on pre-trajectory word alignments (\texttt{simalign\_inter}). Random sampling concentrates mass near $\mathrm{NIR}{=}0$ and higher Spearman $\rho$, reflecting predominantly monotonic alignments. NIR-stratified selection deliberately retains harder reordering buckets (10\% high-NIR quota), so its pooled distribution spreads to larger NIR values even though the policy is monotonicity-aware. Dashed vertical lines in the figure mark the direction-wise means; summary moments appear in the table.

\begin{table}[htbp]
  \centering
  \caption{Direction-wise mean and variance of NIR and Spearman $\rho$ for NIR-stratified vs.\ random selection. NIR moments use $\mathrm{NIR}\le 50\%$; Spearman uses all retained samples.}
  \label{tab:appendix_nir_random_stats}
  \small
  \setlength{\tabcolsep}{3pt}
  \begin{tabular}{@{}llrrrr@{}}
    \toprule
    \textbf{Selection} & \textbf{Dir.} &
    \multicolumn{2}{c}{\textbf{NIR (\%)}} &
    \multicolumn{2}{c}{\textbf{Spearman $\rho$}} \\
    \cmidrule(lr){3-4}\cmidrule(lr){5-6}
    & &
    $\mu$ & $\sigma^2$ & $\mu$ & $\sigma^2$ \\
    \midrule
    NIR stratified & En$\rightarrow$Zh &
    13.79 & 146.64 & 0.747 & 0.103 \\
    NIR stratified & Zh$\rightarrow$En &
    15.86 & 160.69 & 0.687 & 0.147 \\
    Random         & En$\rightarrow$Zh &
    9.66  & 131.46 & 0.821 & 0.081 \\
    Random         & Zh$\rightarrow$En &
    9.92  & 147.47 & 0.760 & 0.176 \\
    \bottomrule
  \end{tabular}
\end{table}

\subsection{ASR-Based Pair Filtering}
\label{app:asr-pair-filtering}

Before paired examples enter S2ST training, we filter synthesized targets with ASR-based content checks. For the trajectory-ready pool, the En$\rightarrow$Zh direction contains 1,470,930 candidates (1,589.62 hours) and the Zh$\rightarrow$En direction contains 1,206,014 candidates (816.06 hours). We apply language-specific error-rate thresholds of 5\% for Chinese and 7\% for English. After filtering, 78.76\% of En$\rightarrow$Zh candidates and 80.19\% of Zh$\rightarrow$En candidates are retained. This step removes content errors and duration mismatches before semantic-code supervision is created. We keep these thresholds deliberately above zero rather than requiring near-perfect ASR match: voice-cloned synthesis legitimately carries source-accent coloring and disfluencies (e.g., filled pauses), and an overly strict threshold would trade this prosodic diversity for ASR-``clean'' but flat speech.

\section{Training Implementation Details and Hyperparameters}
\label{app:training-details}

Table~\ref{tab:training_hparams} summarizes the stage-level training budgets and lists the key hyperparameters for each stage. All runs use eight A800 GPUs. For each run, we select the checkpoint with the lowest held-out evaluation loss on per-dataset validation splits sampled at 0.5\%, capped at 1{,}000 examples each. For both models, the speech encoder and Thinker are initialized from Qwen2.5-Omni~\cite{xu2025qwen25omni}, and the same semantic-code pipeline~\cite{li2025dualcodec}, flow-matching acoustic model~\cite{chen2026flexivoice,le2023voiceboxtextguidedmultilingualuniversal}, and vocoder are adopted.

\begin{table*}[t]
  \centering
  \caption{Key hyperparameters for the main Thinker--Talker line and matched Dec-only baseline. ``Tok/GPU'' denotes tokens per GPU; ``GPU Time'' reports A800 GPU-hours for the corresponding stage.}
  \label{tab:training_hparams}
  \small
  \setlength{\tabcolsep}{4pt}
  \resizebox{\textwidth}{!}{%
  \begin{tabular}{lllcccc}
    \toprule
    \textbf{Line} & \textbf{Stage} & \textbf{Trainable scope} & \textbf{Tok/GPU} & \textbf{GPU Time} & \textbf{LR schedule} & \textbf{Key setting} \\
    \midrule
    \multirow{3}{*}{Thinker--Talker}
      & Stage~1 & -- / Full & 10k & 416 & $4\times10^{-4}\!\rightarrow\!4\times10^{-5}$ & 2 epoch TTS \\
      & Stage~2 & LoRA r32 / Full & 6.4k & 136 & $1\times10^{-4}\!\rightarrow\!2\times10^{-5}$ & multitask mixture \\
      & Stage~3 & LoRA r8 / LoRA r8 & 5.2k & 176 & $3\times10^{-5}\!\rightarrow\!2\times10^{-6}$ & 1 epoch trajectory data \\
    \midrule
    \multirow{3}{*}{Dec-only}
      & Stage~1 & LoRA r32/r32 & 8k / 10k & 1{,}248 & $1\times10^{-4}\!\rightarrow\!1\times10^{-5}$ & 1 epoch S2T\&MT/ 2 epoch TTS \\
      & Stage~2 & LoRA r64 & 6.4k & 112 & $1\times10^{-4}\!\rightarrow\!2\times10^{-5}$ & same multitask mixture \\
      & Stage~3 & LoRA r8 & 5.2k & 136 & $3\times10^{-5}\!\rightarrow\!2\times10^{-6}$ & 1 epoch trajectory data \\
    \bottomrule 
  \end{tabular}%
  }
\end{table*}

\subsection{Thinker--Talker}

\paragraph{Stage~1: Talker warmup.}
We initialize the Talker from the Qwen2.5-Omni talker module (see Appendix~\ref{app:talker-init-control} for the impact of different initialization methods). We then replace the Talker embedding and output head with the DualCodec vocabulary (16{,}384) and add four special tokens: \texttt{codec\_bos}, \texttt{codec\_eos}, \texttt{codec\_pad}, and \texttt{codec\_mask}. We unfreeze the Talker while freezing the Thinker, training the Talker using only TTS data for 2 epochs.

\paragraph{Stage~2: multitask joint training.}
\label{par:talker_stage2}
We then switch to joint multitask training on ASR, S2TT, MT, TTS, and S2ST, with a sampling ratio of ASR:S2TT:MT:TTS:S2ST = $0.2{:}1{:}0.5{:}1{:}1.5$. To mitigate the scarcity of ASR data, we randomly replace the S2TT target with the ASR task with a probability of 10\%. In this stage, we add LoRA modules to the Thinker and jointly train the Thinker's LoRA and the Talker. This aligns the Talker's hidden space while strengthening the Thinker's intelligence. The training target is set to 30k steps.

\paragraph{Stage~3: streaming trajectory finetuning.}
\label{par:talker_stage3}
We merge the previous adapters into the base weights and instantiate fresh low-rank adapters for both the Thinker and Talker. We freeze the embeddings and prediction heads, and continue training on quality-filtered multi-task trajectories (including simultaneous ASR, S2TT, MT, and S2ST) for 1 epoch. In the default main-line setting, latency multipliers are sampled using probabilities that linearly decay from $1$ to $M_{\max}$, with $M_{\max}{=}12$.

\subsection{Dec-only}
\label{app:dec-only}



We first attempted to train the unified decoder with exactly the same data mixture and optimization recipe used by the Talker Stage~2 setup. In practice, this recipe did not allow the Dec-only model to reach comparable performance, due to inherent representational conflicts between multi-modal understanding and generation tasks~\cite{shi2025balancing,wu2024janusdecouplingvisualencoding}. Increasing the proportion of understanding-oriented data improved translation and recognition behavior but weakened speech generation, whereas increasing the proportion of generation-oriented data had the opposite effect. To resolve this optimization conflict under fixed data and compute budgets, we introduced a dual-branch warmup stage---separating understanding-heavy and generation-heavy adaptations before performing parameter fusion, drawing inspiration from separate modality-specialized adapter architectures~\cite{xu2025modality}.

\paragraph{Stage~1: dual-branch LoRA warmup.}
We initialize two different sets of LoRA parameters for the two branches. The understanding branch emphasizes MT, S2TT, ASR, and related tasks, which account for roughly 70\% of its training exposure, and is trained for 1 epoch on all data. The generation branch emphasizes TTS, S2ST, and related generation tasks, which account for roughly 90\% of its exposure, and is trained for 2 epochs, exhausting the TTS data.

\paragraph{Stage~2: multitask joint training.}
We then merge the understanding and generation branches and continue with full multitask training. The weight-fusion procedure is described in Section~\ref{par:weight_fusion}. After fusion, Stage~2 uses the same multitask configuration as the Talker Stage~2 (Section~\ref{par:talker_stage2}).

\paragraph{Stage~3: streaming trajectory finetuning.}
For streaming finetuning, we follow the same Stage~3 trajectory recipe used by the Talker line (Section~\ref{par:talker_stage3}). We first merge the previous-stage adapters into the base model, then instantiate fresh low-rank adapters and freeze the embedding and prediction-head parameters.

\paragraph{Latency multiplier sampling.}
During streaming finetuning, each trajectory is additionally viewed under a sampled latency multiplier $m \in \{1,\ldots,M_{\max}\}$ that controls how aggressively source chunks are merged before emission. The \textsc{Random} and \textsc{NIR} ablations in Table~\ref{tab:realsi_doc_ablation} both draw $m$ uniformly, i.e., $P(m)=1/M_{\max}$, and differ only in whether the trajectory pool is randomly selected or NIR-stratified. The \textsc{NIR+Linear} variant keeps the same NIR-stratified pool but replaces uniform sampling with a truncated linear schedule that front-loads low-latency supervision: 
$$\tilde{w}_m=\max(a-b(m-1),0)$$ 
and $P(m)=\tilde{w}_m/\sum_{k=1}^{M_{\max}}\tilde{w}_k$. We use an intercept of $a{=}0.18$, a slope of $b{=}0.017$, and $M_{\max}{=}12$. The \textsc{NIR+Gaussian} variant also keeps the same NIR-stratified pool, but samples multipliers from a discretized Gaussian weighting centered on lower-latency tiers. We set the Gaussian mean to $3.5$ and the variance to $4.0$, then normalize the resulting weights over $m\in\{1,\ldots,M_{\max}\}$. The main-line Stage~3 recipe also uses uniform multiplier sampling over $\{1,\ldots,M_{\max}\}$.

\paragraph{Initialization fairness.}
Both lines are initialized from the same Qwen2.5-Omni family of publicly released weights: Dec-only inherits the Qwen2.5-Omni LLM body, and Thinker--Talker additionally inherits the Qwen2.5-Omni talker decoder body. In both lines, the codec embedding and output head---the only parameters that interact directly with DualCodec vocabulary---are freshly initialized. Initializing the Talker from Qwen serves as a modest optimization aid rather than the primary source of the performance gain (ablations on the initialization method are provided in Appendix~\ref{app:talker-init-control}). Furthermore, in terms of compute, the Dec-only Stage~1 dual-branch LoRA warmup uses substantially more GPU-hours (Table~\ref{tab:training_hparams}), making the overall comparison conservative with respect to our system.

\paragraph{Weighted fusion of Thinker LoRA.} 
\label{par:weight_fusion}
Some Dec-only runs use branch-focused LoRA adapters that are fused before continued multitask or trajectory finetuning. Given $N$ adapters $\{(A_i,B_i)\}_{i=1}^{N}$ and normalized weights $\{w_i\}_{i=1}^{N}$, we construct a single merged adapter $(\bar A,\bar B)$ that exactly preserves the weighted sum of the individual updates, i.e., $\bar B\bar A = \sum_{i=1}^{N} w_i B_i A_i$. This is achieved by block-concatenating the scaled factors:
\begin{equation}
\begin{aligned}
  \bar A &=
  \begin{bmatrix}
    \sqrt{|w_1|}\,A_1 \\
    \vdots \\
    \sqrt{|w_N|}\,A_N
  \end{bmatrix}, \\
  \bar B &= \bigl[C_1 \ \cdots\ C_N\bigr], \\
  C_i &= \mathrm{sgn}(w_i)\sqrt{|w_i|}\,B_i,
  \qquad i=1,\ldots,N.
\end{aligned}
\end{equation}
The newly merged adapter can then be finetuned further under the standard schedule.

\section{Inference and Decoding}
\label{app:inference}

The offline S2ST task takes full source audio utterances as input and outputs the full target audio. For both architectures, we set \texttt{repetition\_penalty=1.2} for text decoding and \texttt{repetition\_penalty=1.4} for code decoding.

Streaming evaluation uses finite source context and latency tiers induced by trajectory multipliers. Following InfiniSST~\cite{ouyang2025infinisst}, we preserve system prompts and keep a bounded history window across turns. InfiniSST implements this strategy with a sinking-cache mechanism built on attention sinks that retain pre-RoPE key-value states. In contrast, we directly preserve the dynamic key-value cache from the previous turn. Once the history window limit is reached, we retain the latest $\lfloor 16/m \rfloor$ history turns and perform a full forward pass to obtain a new KV cache, where $m$ is the latency multiplier.

For both architectures, streaming text decoding uses \texttt{no\_repeat\_ngram\_size=5} and \texttt{repetition\_penalty=1.2}. Beam search is additionally applied where explicitly mentioned. Semantic-code decoding uses greedy decoding with \texttt{repetition\_penalty=1.4} and \texttt{no\_repeat\_ngram\_size=5}, following the repetition-suppression strategy used in InfiniSST. These decoding settings are kept consistent between the Talker and Dec-only models.

Directly decoding a mel spectrogram from a single chunk with the flow-matching model and vocoder introduces audible discontinuities across chunk boundaries and produces unstable prosody and rhythm in long-form audio. To reduce these boundary artifacts, we preserve the latest 4\,s of source audio and the latest 2\,s of generated audio as prompt context for the flow-matching model. This improves timbre consistency and smooths transitions across chunks.

\section{Evaluation Protocols}
\label{app:offline-eval-details}

\subsection{Offline TTS Evaluation}

\paragraph{Evaluation datasets.}
\textbf{Seed-TTS}
Supplementary zero-shot TTS evaluation uses English and Chinese test sets following the Seed-TTS protocol~\cite{anastassiou2024seedttsfamilyhighqualityversatile}. These diagnostics are used only to assess speech-generation quality apart from the main S2ST benchmarks.

\paragraph{Evaluation metrics.}
\begin{itemize}
  \item \textbf{WER} measures content consistency between synthesized speech and the target text. English recognition uses Whisper-Large-V3~\cite{radford2022robustspeechrecognitionlargescale}, while Chinese recognition uses Paraformer~\cite{paraformer}.
  \item \textbf{SIM-O} measures speaker similarity using WavLM-Large~\cite{wavlm} speaker embeddings and cosine similarity against the reference target speech.
\end{itemize}

\subsection{Offline Speech Understanding Evaluation}

\paragraph{Evaluation datasets.}
\textbf{LibriSpeech (English ASR).}
We report English speech understanding on LibriSpeech \textit{test-clean}~\cite{panayotov2015librispeech}.

\textbf{WenetSpeech (Mandarin ASR).}
We report Mandarin speech understanding on the WenetSpeech TestNet split~\cite{zhang2022wenetspeech10000hoursmultidomain}.

\textbf{CoVoST~2 (offline S2TT).}
We report offline speech-to-text translation on the En$\rightarrow$Zh and Zh$\rightarrow$En directions of CoVoST~2~\cite{wang2021covost}.

\paragraph{Evaluation metrics.}
\begin{itemize}
  \item \textbf{WER / CER} are used for English and Mandarin ASR respectively. English ASR uses the LibriSpeech evaluation protocol, and Mandarin ASR uses the WenetSpeech evaluation protocol.
  \item \textbf{BLEU} is used for offline S2TT on CoVoST~2 under the same text normalization and SacreBLEU~\cite{post-2018-call} setup described below for translation quality metrics.
\end{itemize}

\subsection{Offline and Streaming S2ST Evaluation Details}
\label{app:baseline}

\paragraph{Evaluation datasets.}
\textbf{CVSS-T (offline S2ST).}
We use the Chinese (ZH) and English (EN) test sets from CVSS-T~\cite{jia-etal-2022-cvss}, which contain 4,897 utterance pairs (8.2 hours for Chinese and 6.3 hours for English). Both En$\rightarrow$Zh and Zh$\rightarrow$En directions are evaluated.

\textbf{RealSI (streaming S2TT and S2ST).}
RealSI~\cite{cheng2025seedliveinterpret20endtoend} is a long-form Chinese--English simultaneous interpretation benchmark comprising natural, casual speech across ten domains. It provides sentence-level and long-form evaluation settings. The sentence-level setting contains 51:12 minutes / 431 segments for En$\rightarrow$Zh and 44:17 minutes / 347 segments for Zh$\rightarrow$En.

\textbf{ACL60/60-dev (long-form S2TT).}
We additionally evaluate long-form En$\rightarrow$Zh streaming S2TT on ACL60/60-dev~\cite{salesky2023evaluating}, which consists of 5 real-world ACL 2022 oral-presentation talks in English, containing roughly one hour of audio in total under the IWSLT realistic-condition setting.

\paragraph{Evaluation inputs and references.}
Each evaluation example contains a reference translation text, a reference speech waveform used for speech-side comparison, and the model-generated target speech. Text-based metrics compare model text or ASR transcripts against the target reference. Speech similarity metrics compare the generated waveform against the paired reference waveform provided by the evaluation manifest.

\paragraph{Translation quality metrics.}
Unless otherwise stated, our offline S2ST comparison follows the UniSS evaluation protocol so that CVSS-T results remain directly comparable with recent offline S2ST systems.
\begin{itemize}
  \item \textbf{Text-BLEU}\label{metric:text-bleu} evaluates generated translation text with \textit{corpus-level} SacreBLEU~\cite{post-2018-call} following UniSS preprocessing. English text is lowercased and stripped of punctuation while preserving apostrophes. Chinese text is converted to simplified Chinese, punctuation is removed, and characters are separated by spaces. We use SacreBLEU's \texttt{zh} tokenizer for Chinese and \texttt{13a} for English.
  \item \textbf{ASR-BLEU}\label{metric:asr-bleu} evaluates whether the synthesized speech preserves translation content. Generated speech is transcribed with Whisper-Large-V3~\cite{radford2022robustspeechrecognitionlargescale} for English and Paraformer-zh~\cite{paraformer} for Chinese, then scored against the target reference under the same normalization used for Text-BLEU.
\end{itemize}

\paragraph{Speech similarity metrics.}
\begin{itemize}
  \item \textbf{SIM-O} is computed with WavLM-Large~\cite{wavlm} speaker embeddings and cosine similarity against the paired real reference target speech.
  \item \textbf{A.PCP} (AutoPCP) follows the automatic prosodic-consistency protocol used in expressive multilingual S2ST evaluation. We use the Stopes~\cite{andrews-etal-2022-stopes} implementation on paired 16\,kHz waveforms with the released \textbf{AutoPCP-multilingual-v2} comparator on hidden states from layer~9 of a frozen \textbf{Wav2Vec2-Large-XLSR-53} encoder to compare the output against the real source audio.
\end{itemize}

\paragraph{Streaming metrics.}
Streaming experiments use SimulEval-style latency metrics together with task-specific quality metrics.
\begin{itemize}
  \item \textbf{Sentence-level S2TT/S2ST} uses LAAL~\cite{papi2022over} and computation-aware LAAL (LAAL\_CA). Sentence-level S2TT quality is reported with SimulEval text BLEU; sentence-level S2ST quality is reported with ASR-BLEU, A.PCP, and SIM-O from the generated speech.
  \item \textbf{Long-form S2TT} uses StreamLAAL~\cite{papi2024streamatt} and StreamLAAL\_CA. The main trade-off plots use the stream-side BLEU emitted by the StreamLAAL evaluator so that quality and latency come from the same long-form run.
  \item \textbf{Long-form S2ST} supplementary results report ASR-BLEU on the generated speech using corpus-level SacreBLEU.
  \item \textbf{RTF} is reported for sentence-level S2ST systems with instance logs, using active compute time divided by source-speech duration.
\end{itemize}

\section{Supplementary Evaluation Results}

\subsection{Offline Speech Understanding and Translation}
\label{app:offline-mt-asr}

Table~\ref{tab:offline_mt_asr} compares public spoken-language models with our Dec-only baseline and Talker main line on offline speech understanding and speech-to-text translation. We report English ASR on LibriSpeech \textit{test-clean}, Mandarin ASR on WenetSpeech TestNet, and CoVoST~2 En$\rightarrow$Zh / Zh$\rightarrow$En BLEU under the preprocessing described in Section~\ref{app:offline-eval-details}. Published baseline numbers are taken from the original papers when available; the public comparison set includes SALMONN~\cite{anonymous2023salmonn}, GLM-4-Voice-Base~\cite{zeng2024glm4voice}, MinMo~\cite{chen2025minmomultimodallargelanguage}, Qwen-Omni2.5~\cite{xu2025qwen25omni}, Step-Audio~2~\cite{wu2025stepaudio2technicalreport}, Qwen3-Omni~\cite{xu2025qwen3omnitechnicalreport}, and Covo-Audio~\cite{wang2026covoaudiotechnicalreport}. 
This table isolates understanding and translation quality from speech-generation behavior, which is evaluated separately on SEED in the next subsection.

\begin{table}[htbp]
  \centering
  \caption{Offline speech understanding and speech-to-text translation. ASR columns report WER/CER (lower is better); S2TT columns report BLEU (higher is better). \textbf{Bold} marks the best result in each column.}
  \label{tab:offline_mt_asr}
  \small
  \setlength{\tabcolsep}{3pt}
  \begin{tabular}{@{}llcccc@{}}
    \toprule
    & & \multicolumn{2}{c}{\textbf{ASR}} & \multicolumn{2}{c}{\textbf{S2TT}} \\
    \cmidrule(lr){3-4}\cmidrule(lr){5-6}
    \textbf{Model} & \textbf{\#} & En & Zh & En$\rightarrow$Zh & Zh$\rightarrow$En \\
    \midrule
    SALMONN & 7B & 2.10 & --- & 33.10 & --- \\
    GLM-4-Voice-Base & 9B & 2.82 & --- & --- & --- \\
    Minmo & 7B & 1.70 & 6.80 & 46.70 & 26.00 \\
    Qwen-Omni2.5 & 7B & 1.80 & 5.90 & 41.40 & 29.40 \\
    Step-Audio 2 & 30B & \textbf{1.17} & \textbf{4.67} & 49.01 & \textbf{29.51} \\
    Qwen3-Omni & 30A3 & 1.22 & 4.69 & 48.72 & 21.50 \\
    Covo-Audio & 7B & 1.45 & 7.23 & \textbf{49.84} & 26.77 \\
    \midrule
    \textbf{Ours (Dec-only)} & 3B & 2.10 & 7.43 & 45.70 & 24.65 \\
    \textbf{Ours (Talker)} & 3B & 2.78 & 6.79 & 45.09 & 26.68 \\
    \bottomrule
  \end{tabular}
\end{table}

Overall, the two architectures remain in a similar offline regime relative to public SLMs, suggesting that introducing an independent Talker does not materially degrade the shared thinker's speech understanding or translation ability. The observed differences are modest and task-dependent rather than systematic, so we treat offline understanding as preserved and focus on architecture-specific speech-generation behavior on SEED below.

\subsection{Zero-Shot Speech Generation on SEED}

We evaluate the zero-shot speech-generation ability of our models on the Seed-TTS benchmark. All TTS results use the in-context-learning (ICL) protocol on SEED, focusing on content consistency (WER) and speaker similarity (SIM-O). Table~\ref{tab:zero_shot_speech_generation_table} compares public models and our two structures. The public TTS and speech-language baselines include MaskGCT~\cite{maskgct}, E2~TTS~\cite{eskimez2024e2ttsembarrassinglyeasy}, F5-TTS~\cite{f5tts}, CosyVoice~2~\cite{du2024cosyvoice2scalablestreaming}, GLM-4-Voice-Base~\cite{zeng2024glm4voice}, MinMo~\cite{chen2025minmomultimodallargelanguage}, and Qwen-Omni2.5~\cite{xu2025qwen25omni}.

\begin{table}[htbp]
  \centering
  \caption{Zero-shot speech generation on SEED. WER$\downarrow$ and SIM-O$\uparrow$ are reported for the English and Chinese subsets; \textbf{bold} marks the best result in each column. Missing entries are not reported in the original paper.}
  \label{tab:zero_shot_speech_generation_table}
  \small
  \setlength{\tabcolsep}{3pt}
  \resizebox{\columnwidth}{!}{%
  \begin{tabular}{@{}lcccc@{}}
    \toprule
    & \multicolumn{2}{c}{\textbf{SEED EN}} & \multicolumn{2}{c}{\textbf{SEED ZH}} \\
    \cmidrule(lr){2-3}\cmidrule(lr){4-5}
    \textbf{Model} & WER$\downarrow$ & SIM-O$\uparrow$ & WER$\downarrow$ & SIM-O$\uparrow$ \\
    \midrule
    MaskGCT & 2.62 & 0.714 & 2.27 & \textbf{0.774} \\
    E2 TTS & 2.19 & 0.710 & 1.97 & 0.730 \\
    F5-TTS & \textbf{1.83} & 0.647 & \textbf{1.56} & 0.741 \\
    CosyVoice 2 & 2.57 & 0.652 & 1.45 & 0.748 \\
    \midrule
    GLM-4-Voice-Base 9B & 2.91 & --- & 2.10 & --- \\
    Minmo 7B & 2.90 & --- & 2.48 & --- \\
    Qwen-Omni2.5 7B & 2.72 & \textbf{0.747} & 1.70 & 0.752 \\
    \midrule
    \textbf{Ours (Dec-only)} & 2.73 & 0.665 & 2.34 & 0.741 \\
    \textbf{Ours (Talker)} & 1.93 & 0.652 & 1.87 & 0.723 \\
    \bottomrule
  \end{tabular}
  }
\end{table}

The SEED comparison shows the same structural pattern as the offline and streaming S2ST results: separating code prediction from the unified decoder mainly improves content consistency, while speaker similarity remains in a comparable range. This serves as supplementary evidence that the Talker primarily helps the model say the correct content more reliably.

\paragraph{Initialization control.}
\label{app:talker-init-control}
To test whether initialization materially affects TTS performance, we run a matched TTS-only control on Emilia and evaluate on SeedTTS after one epoch. We compare a Talker initialized from the reused Qwen talker body, a Talker initialized from scratch, all with the Thinker frozen, and a Dec-only baseline with a TTS LoRA branch. Table~\ref{tab:talker_init_seedtts} shows that initialization has only a minor effect within the Talker architecture: Qwen initialization provides a small advantage, but the from-scratch Talker remains close and still outperforms the Dec-only TTS branch. We therefore treat Qwen initialization as a modest optimization aid rather than the main source of the gain, which is better explained by the architectural separation itself than by training duration or initialization alone.

\begin{table}[htbp]
  \centering
  \caption{SeedTTS TTS-only initialization control. Lower WER is better; higher SIM-O is better. Both rows use one-epoch TTS-only warmup on Emilia.}
  \label{tab:talker_init_seedtts}
  \small
  \setlength{\tabcolsep}{3pt}
  \resizebox{\columnwidth}{!}{%
  \begin{tabular}{@{}lcccc@{}}
    \toprule
    & \multicolumn{2}{c}{\textbf{SEED EN}} & \multicolumn{2}{c}{\textbf{SEED ZH}} \\
    \cmidrule(lr){2-3}\cmidrule(lr){4-5}
    \textbf{Model} & WER$\downarrow$ & SIM-O$\uparrow$ & WER$\downarrow$ & SIM-O$\uparrow$ \\
    \midrule
    \textbf{Talker (Qwen-initialized)} & 2.03 & 0.647 & 1.80 & 0.723 \\
    \textbf{Talker (From-scratch)} & 2.41 & 0.643 & 1.85 & 0.723 \\
    \textbf{Dec-only (TTS-LoRA)} & 3.98 & 0.657 & 3.22 & 0.738 \\
    \bottomrule
  \end{tabular}
  }
\end{table}

\subsection{RealSI Long-form S2TT Trade-off}
\label{app:realsi_doc_s2tt}

\begin{figure*}[htbp]
  \centering
  \includegraphics[width=0.92\linewidth]{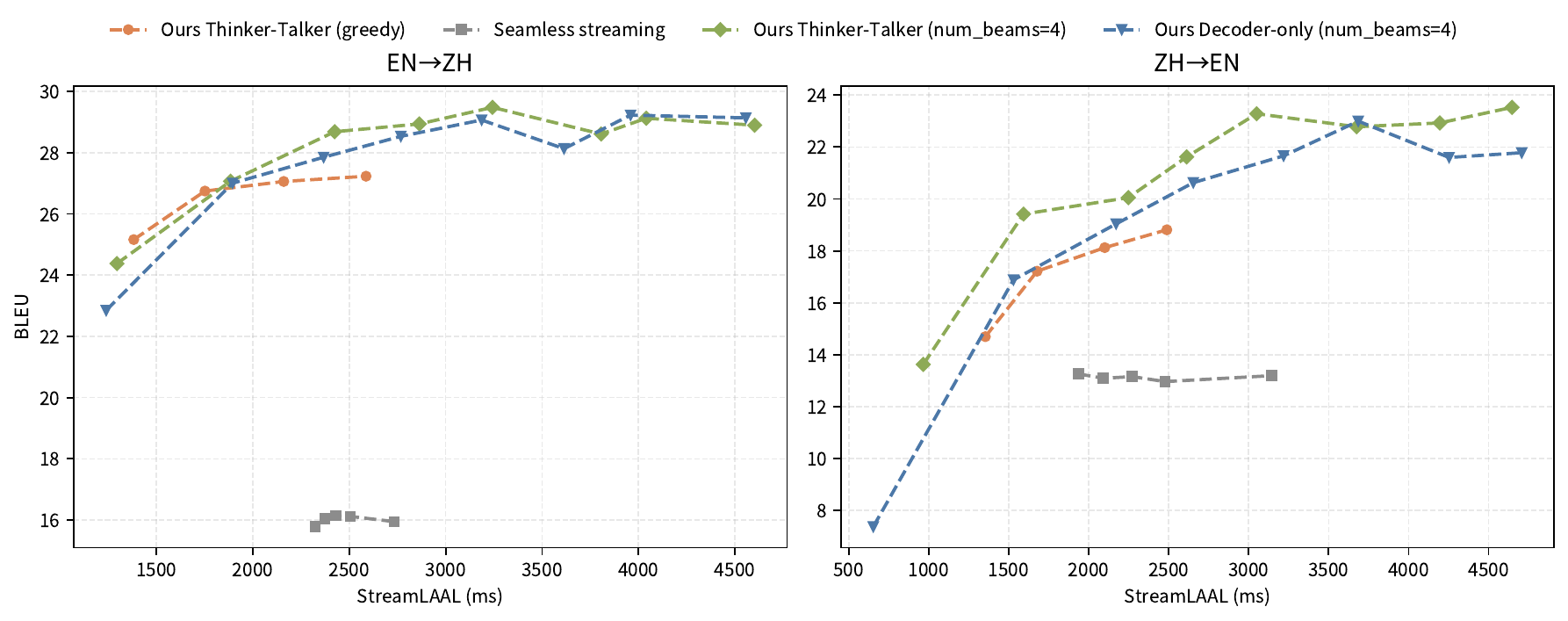}
  \caption{RealSI long-form streaming S2TT trade-off for both directions.}
  \label{fig:realsi_document_tradeoff_appendix}
\end{figure*}

The main text anchors our long-form performance claims using RealSI long-form S2ST against LiveInterpret~2.0 (Table~\ref{tab:realsi_streaming_s2st_asr_bleu}) and ACL60/60 against academic systems. For completeness, Figure~\ref{fig:realsi_document_tradeoff_appendix} reports the RealSI long-form streaming S2TT trade-off, with Seamless streaming as an additional reference. Consistent with our S2ST findings, the Thinker--Talker model forms a strong quality--latency frontier above the external baseline. On both En$\rightarrow$Zh and Zh$\rightarrow$En, our beam-search Talker (\textsc{nb4}) substantially outperforms Seamless streaming, maintaining leads of roughly 10 to 13 BLEU across all comparable latency tiers.

Crucially, this evaluation reaffirms the structural advantage of the two-stream Talker over the unified decoder for long-form continuous generation. Across both directions, the Talker strictly dominates the matched Dec-only baseline. This architectural advantage is most pronounced on the challenging Zh$\rightarrow$En direction at ultra-low latencies, where the unified Dec-only model suffers a catastrophic collapse (plummeting to roughly 7.5 BLEU), whereas the Talker maintains robust translation fidelity. Furthermore, while greedy decoding provides a fast, competitive baseline at low latencies, transitioning to beam search (\textsc{nb4}) yields a massive and necessary quality boost to sustain optimal long-form generation as the read window relaxes.

\section{Computation-Aware Simultaneous Evaluation}
\label{app:ca-eval}

This subsection examines whether our main latency--quality conclusions remain stable after accounting for active inference cost. We reuse the same RealSI runs as in the main paper and evaluate them with three complementary views: (i)~LAAL\_CA / StreamLAAL\_CA, which augment standard latency metrics with computation-aware delay; and (ii)~active compute RTF, defined as the total non-idle inference time divided by source duration. These metrics are intended to separate listener-side latency from pure wall-clock computation.

\begin{figure*}[h]
  \centering
  \includegraphics[width=0.92\linewidth]{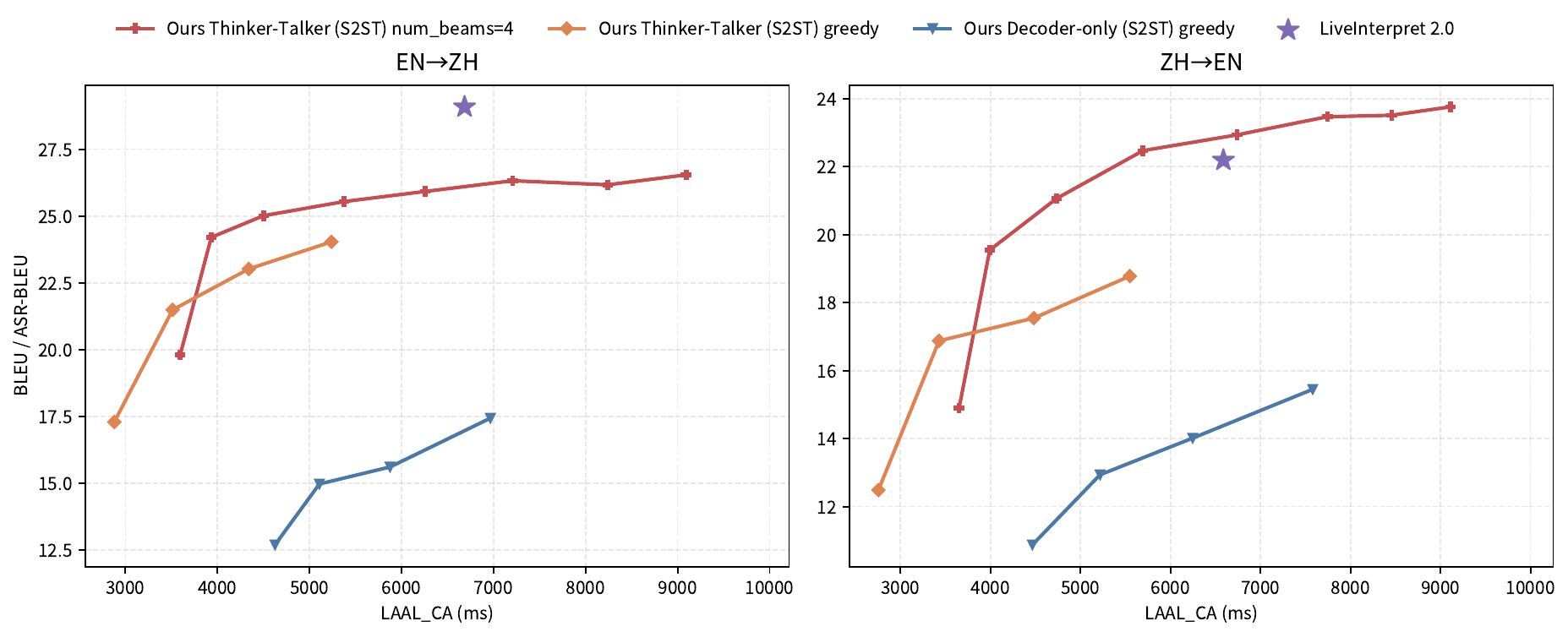}
  \caption{RealSI Sentence-level simultaneous S2ST under LAAL\_CA. ASR-BLEU is plotted against computation-aware latency on RealSI sentence test sets.}
  \label{fig:appendix_sentence_ca}
\end{figure*}

\subsection{Sentence-level S2ST LAAL\_CA.}
Figure~\ref{fig:appendix_sentence_ca} replots the sentence-level S2ST frontier, shifting the horizontal axis to computation-aware latency (LAAL\_CA). Introducing computation awareness adds an average delay of approximately 2 seconds across all tiers compared to the ideal, non-CA latency. Notably, the latency gap on the x-axis between our beam-search (\textsc{nb4}) and greedy Talker configurations is remarkably small. This indicates that the frozen flow-matching acoustic backend consumes the vast majority of the wall-clock processing time, making the translation quality gains of beam search highly worthwhile. 

Furthermore, the \textsc{Dec-only} structure exhibits a severe latency penalty, shifting significantly to the right. Predicting dense acoustic codes directly from a single 3B-parameter backbone creates a major computational bottleneck, severely inflating its delay compared with the lightweight 0.4B Talker. Finally, since LiveInterpret~2.0's own paper does not report a computation-aware latency, we re-implement its official API request as a SimulEval agent and measure its LAAL\_CA under the same wall-clock instrumentation as our own systems. Under this measurement, accounting for computation awareness shifts LiveInterpret~2.0's own operating point considerably further right as well, into a latency regime comparable to our $m4$--$m6$ tiers rather than the low-latency end of the frontier. Within this comparable regime, our system is competitive: on Zh$\rightarrow$En it matches or slightly exceeds LiveInterpret~2.0 at an equal or lower computation-aware latency, while on En$\rightarrow$Zh a moderate quality gap of roughly 3 ASR-BLEU points remains at comparable latency.

\begin{figure*}[h]
  \centering
  \includegraphics[width=0.92\linewidth]{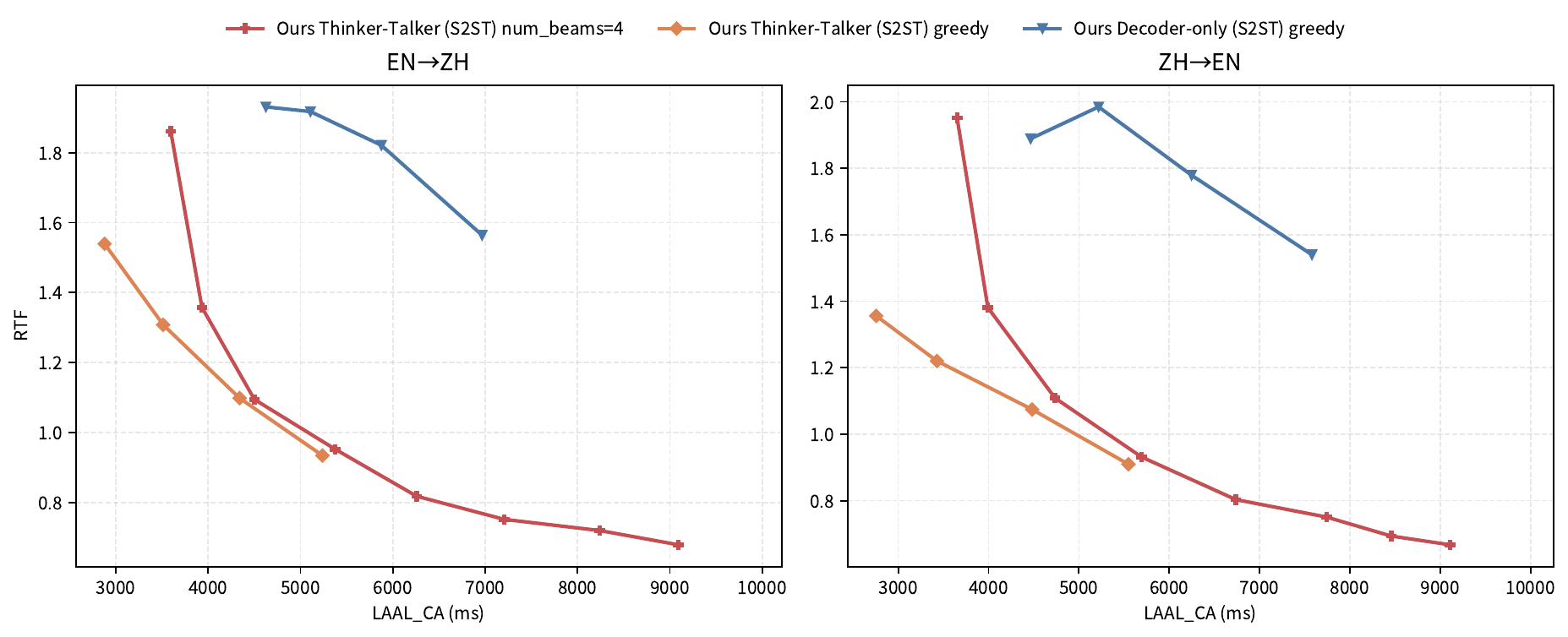}
  \caption{RealSI sentence-level simultaneous S2ST: active compute RTF vs.\ LAAL\_CA. RTF counts only non-idle inference time per source second.}
  \label{fig:appendix_sentence_rtf}
\end{figure*}

\subsection{Sentence-level S2ST Compute RTF}
Figure~\ref{fig:appendix_sentence_rtf} reports active compute RTF against LAAL\_CA. RTF is computed as $\mathrm{RTF}=\sum_i (e_i-\max(d_i,e_{i-1}))/T_{\text{src}}$, where $d_i$ and $e_i$ are the per-chunk input-ready and completion times. Unlike LAAL\_CA, which measures listener delay, this isolates how much active wall-clock compute is required per second of source speech.

Two major trends emerge that align with our previous architectural conclusions. First, the \textsc{Dec-only} structure's RTF is substantially higher than the Talker's at every comparable tier, confirming that the unified decoder is too heavy for efficient continuous audio generation. Second, decoding strategy heavily influences compute at the extremes: the greedy Talker substantially reduces RTF at low latencies compared with \textsc{nb4} (e.g., 1.54 vs.\ 1.86 RTF at En$\rightarrow$Zh $m1$). However, as the latency tier increases, the RTF of greedy and \textsc{nb4} rapidly converge. This reveals that the computational overhead of beam search is primarily driven by the high frequency of chunk-level decisions required at ultra-low latencies. Once chunks are larger (from $m4$ onward), the system becomes usable in real scenarios because the duration of generated audio can cover the compute overhead (RTF ${<}1$).

\begin{figure*}[h]
  \centering
  \includegraphics[width=0.92\linewidth]{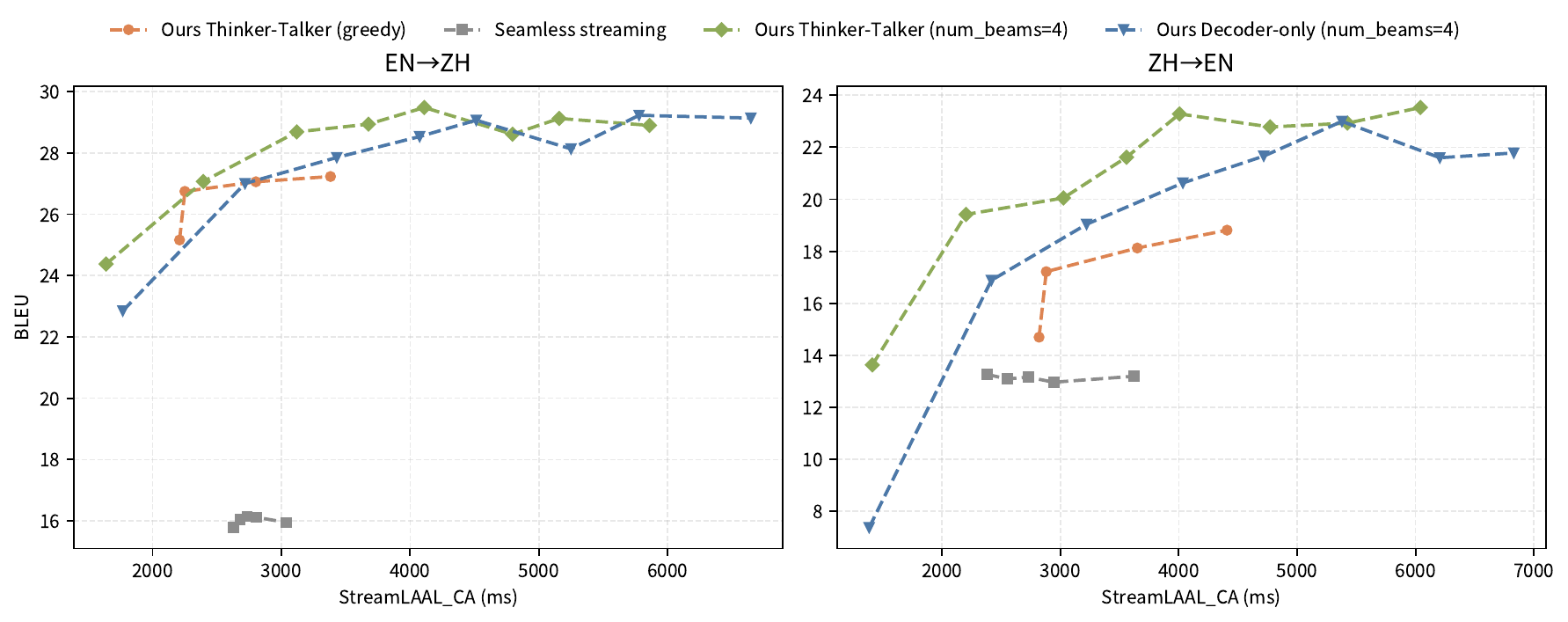}
  \caption{RealSI long-form simultaneous S2TT under StreamLAAL\_CA.}
  \label{fig:appendix_document_ca}
\end{figure*}

\subsection{RealSI Long-form StreamLAAL\_CA}
Figure~\ref{fig:appendix_document_ca} demonstrates that long-form performance rankings remain stable when accounting for active computation time (StreamLAAL\_CA). Across both directions, the Thinker--Talker model (\textsc{nb4}) maintains a large quality advantage over Seamless streaming at every comparable tier. Crucially, the structural advantage of the two-stream architecture persists in long-form generation: the Talker (\textsc{nb4}) consistently yields higher translation fidelity at lower computation-aware latencies than the unified Dec-only baseline. This confirms that the Talker's superiority is a fundamental architectural benefit, not an artifact of a specific latency definition.

\begin{figure}[h]
  \centering
  \includegraphics[width=0.92\linewidth]{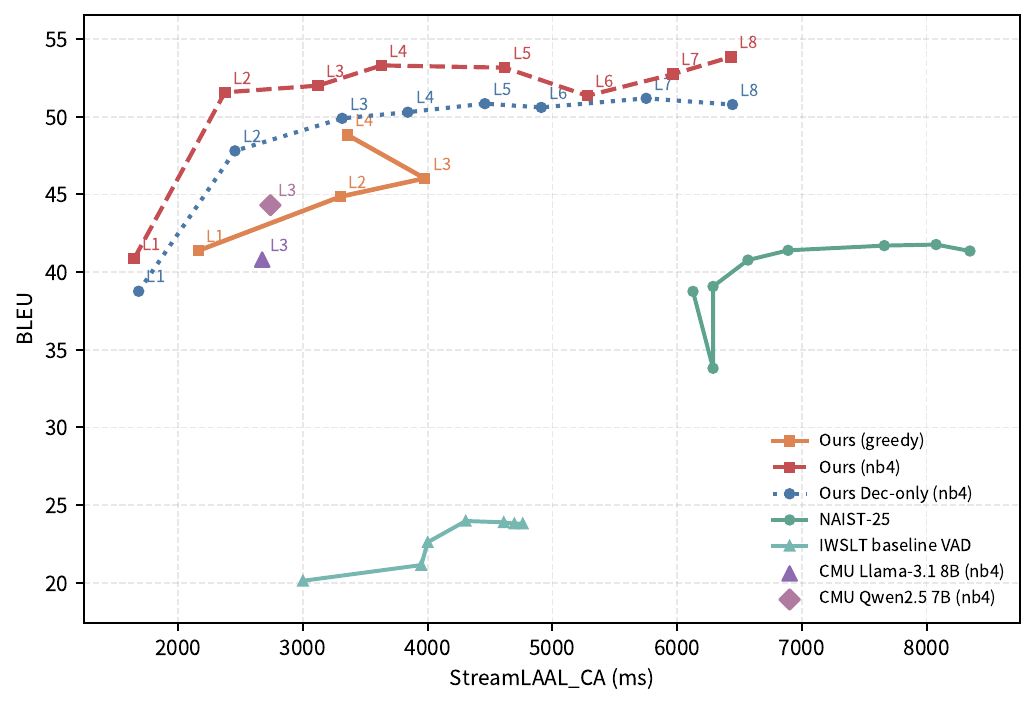}
  \caption{ACL60/60-dev Long-form simultaneous S2TT under StreamLAAL\_CA.}
  \label{fig:appendix_acl_ca}
\end{figure}

\subsection{ACL60/60-dev Long-form StreamLAAL\_CA}
Figure~\ref{fig:appendix_acl_ca} plots StreamLAAL\_CA on ACL60/60-dev, confirming that our system retains its state-of-the-art status even under strict computation-aware evaluation. The Talker (\textsc{nb4}) continues to strictly dominate the Dec-only baseline and vastly outperforms external systems such as NAIST-25 and the IWSLT baseline. Most notably, our computationally inexpensive greedy Talker configuration achieves a quality--latency trade-off highly competitive with CMU's heavier 7B and 8B models operating under beam search. This proves that our trajectory-and-Talker recipe provides a highly compute-efficient, state-of-the-art frontier for unbounded streaming S2TT.

\section{Reproducibility and ASR Sensitivity}
\label{app:asr-sensitivity}

\subsection{Training-Seed Reproducibility}
\label{app:seed-reproducibility}

Our main results report single training runs. Rerunning the full-budget main setting across multiple seeds is prohibitively expensive at our training scale, so we instead pick the cheaper 10\% w/o aux paired-S2ST budget ablation (the setting used in Table~\ref{tab:realsi_s2s_budget_ablation}) as a representative proxy: to confirm the reported Talker--Dec-only gaps are not an artifact of random seed noise, we rerun this ablation for both models, in both directions, across 3 random seeds.

\begin{table}[h]
  \centering
  \caption{Training-seed reproducibility of the 10\% w/o aux paired-S2ST budget ablation. Values are ASR-BLEU, mean $\pm$ std over 3 training seeds.}
  \label{tab:seed_reproducibility}
  \small
  \setlength{\tabcolsep}{4pt}
  \resizebox{\columnwidth}{!}{%
  \begin{tabular}{llcccc}
    \toprule
    \textbf{Direction} & \textbf{Model} & \textbf{m1} & \textbf{m2} & \textbf{m3} & \textbf{m4} \\
    \midrule
    \multirow{2}{*}{En$\rightarrow$Zh}
      & Talker & 9.04$\pm$0.54 & 18.67$\pm$0.19 & 20.90$\pm$0.20 & 21.99$\pm$0.16 \\
      & Dec-only & 12.83$\pm$0.12 & 14.16$\pm$0.20 & 14.96$\pm$0.64 & 16.40$\pm$0.42 \\
    \midrule
    \multirow{2}{*}{Zh$\rightarrow$En}
      & Talker & 8.88$\pm$0.38 & 14.89$\pm$0.25 & 17.31$\pm$0.41 & 18.41$\pm$0.48 \\
      & Dec-only & 10.34$\pm$0.68 & 12.34$\pm$0.62 & 13.90$\pm$0.22 & 15.24$\pm$0.18 \\
    \bottomrule
  \end{tabular}%
  }
\end{table}

Table~\ref{tab:seed_reproducibility} shows that the seed std is small ($\le$0.68 BLEU) across all tiers, and these 3-seed averages agree closely with the single-run numbers already reported in Table~\ref{tab:realsi_s2s_budget_ablation}, confirming that our training is stable rather than driven by random seed noise. The Talker--Dec-only gaps at $m2$--$m4$ (up to $\sim$4.5 BLEU) far exceed this variance, so they reflect genuine architectural differences rather than statistical fluctuation.

\subsection{ASR Backend Sensitivity}
\label{app:asr-backend-sensitivity}

ASR-BLEU transcribes synthesized target speech with an ASR system and scores the transcript against the reference translation, so the metric can inherit systematic biases of the chosen ASR backend, particularly when the generated speech carries accents, prosodic variation, or synthesis artifacts. The main paper reports ASR-BLEU with Whisper-large-v3 (English) and Paraformer-zh (Chinese), following Seed-TTS-eval~\cite{anastassiou2024seedttsfamilyhighqualityversatile} and recent S2ST work~\cite{cheng2025uniss,le2025simulmega}. To probe this potential bias, we re-transcribe the identical generated audio---both ours and LiveInterpret~2.0's---with another strong ASR system, Qwen3-ASR-1.7B, and recompute ASR-BLEU under an otherwise identical protocol, at the same $m2$--$m6$ operating points as Table~\ref{tab:realsi_streaming_s2st_asr_bleu}. Because LiveInterpret~2.0 is only accessible through its official API, we query it 3 times and report both LiveInterpret columns below as the mean $\pm$ std over these runs, transcribed with Qwen3-ASR-1.7B and with the original Whisper-large-v3 / Paraformer-zh backend, respectively.

\begin{table}[h]
  \centering
  \caption{\textbf{RealSI streaming S2ST ASR-BLEU under an alternative ASR backend} (Qwen3-ASR-1.7B), compared against the main-paper backend (Whisper-large-v3 / Paraformer-zh) in Table~\ref{tab:realsi_streaming_s2st_asr_bleu}. \textbf{Bold} marks the highest score per row.}
  \label{tab:asr_sensitivity_qwen}
  \scriptsize
  \setlength{\tabcolsep}{3pt}
  \resizebox{\columnwidth}{!}{%
  \begin{tabular}{llccc>{\columncolor{black!10}}c c >{\columncolor{black!10}}c>{\columncolor{black!10}}c}
    \toprule
    & & \multicolumn{5}{c}{\textbf{Ours (Qwen3-ASR)}} & \multicolumn{2}{c}{\textbf{LiveInterpret}} \\
    \cmidrule(lr){3-7} \cmidrule(lr){8-9}
    \textbf{Type} & \textbf{Direction} & \textbf{m2} & \textbf{m3} & \textbf{m4} & \textbf{m5} & \textbf{m6} & \textbf{Qwen3-ASR} & \textbf{Whisper (orig.)} \\
    \midrule
    \multirow{2}{*}{Sent}
      & En$\rightarrow$Zh & 25.47 & 26.40 & 26.70 & 27.30 & 27.35 & \textbf{29.38$\pm$0.36} & 28.70$\pm$0.37 \\
      & Zh$\rightarrow$En & 19.84 & 20.93 & 22.28 & 23.06 & \textbf{23.26} & 22.34$\pm$0.20 & 22.22$\pm$0.14 \\
    \midrule
    \multirow{2}{*}{Doc}
      & En$\rightarrow$Zh & 28.35 & 30.30 & 31.01 & \textbf{31.43} & 31.13 & 30.59$\pm$0.71 & 29.99$\pm$0.72 \\
      & Zh$\rightarrow$En & 22.25 & 22.56 & 24.06 & 26.55 & 25.51 & \textbf{28.03$\pm$0.30} & 26.32$\pm$0.42 \\
    \bottomrule
  \end{tabular}%
  }
\end{table}

As shown in Table~\ref{tab:asr_sensitivity_qwen}, the relative ranking is largely stable across the two backends: in 3 of 4 settings the comparison is unchanged (LiveInterpret ahead on Sent En$\rightarrow$Zh, ours ahead on Sent Zh$\rightarrow$En and Doc En$\rightarrow$Zh). The only backend-sensitive case is long-form Zh$\rightarrow$En, where our score barely moves while LiveInterpret's rises by roughly 1.7 BLEU (26.32 $\rightarrow$ 28.03), widening its lead in that single setting. Overall, the ASR backend shifts absolute scores somewhat without changing the paper's main conclusions.

\section{Human Evaluation and LLM-as-Judge Assessment}
\label{app:human-eval}

Automatic metrics such as ASR-BLEU can be biased by the ASR backend used for transcription. We therefore complement our automatic evaluation with a human listening study and an LLM-as-judge adequacy protocol (VIP), both on RealSI sentence-level samples against LiveInterpret~2.0 and Seamless-Streaming.

\subsection{Human Listening Study}
\label{app:human-mos}

\begin{figure*}[htbp]
  \centering
  \includegraphics[width=0.95\textwidth]{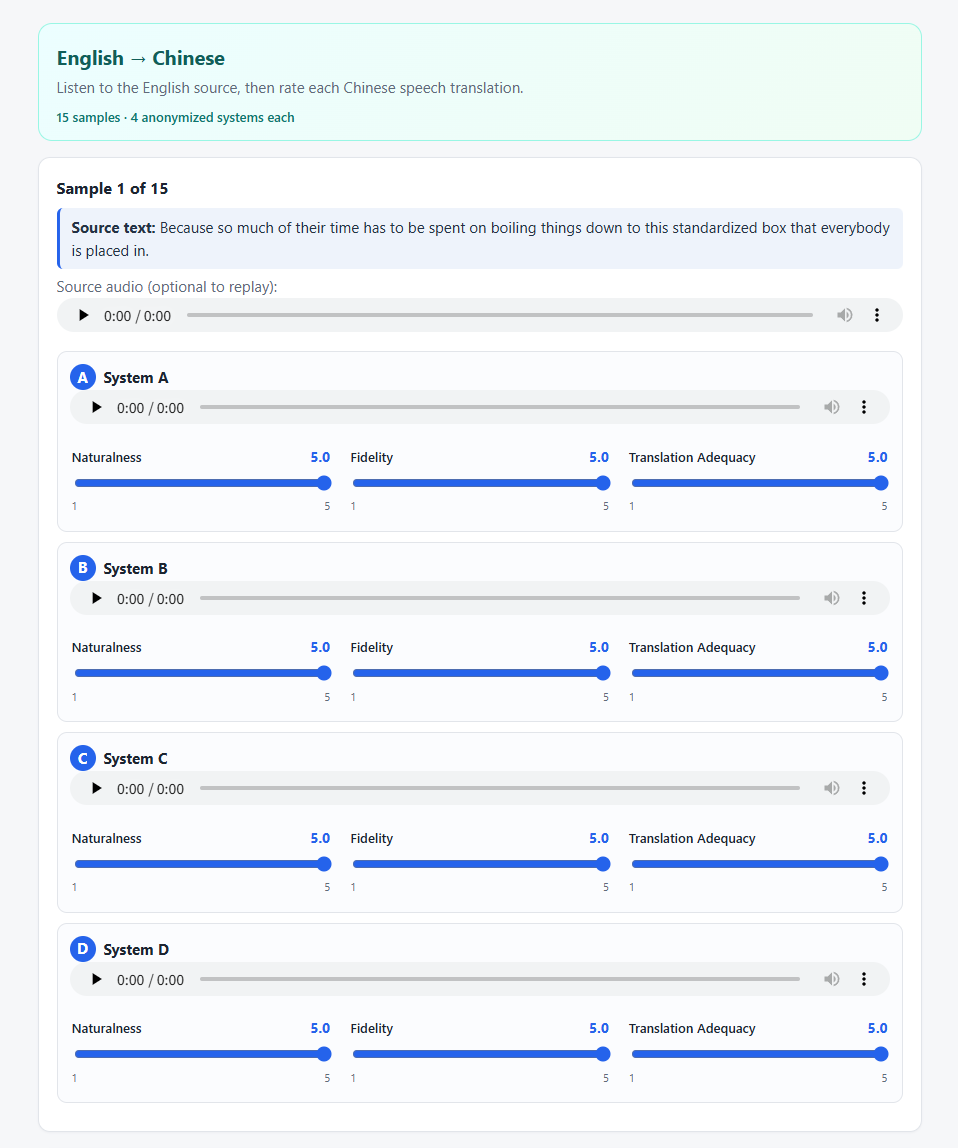}
  \caption{Anonymized rating interface used in the human listening study. Systems are randomly assigned to slots \textsc{A--D} and re-shuffled per sample so that raters cannot infer system identity; raters score Naturalness, Fidelity, and Translation Adequacy for each slot on a 1--5 slider.}
  \label{fig:human_eval_interface}
\end{figure*}

\paragraph{Protocol.} Five evaluators recruited from among the paper authors and other members of the authors' laboratory rate a sampled subset of RealSI sentence-level examples ($N{=}15$ per direction). Participation was voluntary and uncompensated, and no external crowdsourcing platform was used. All evaluators were informed that their ratings would be used for research and publication and voluntarily consented to participate; only aggregated results are reported. Evaluators rate each example along three dimensions on a 1--5 scale, with Overall defined as the mean of the three:
\begin{itemize}
  \item \textbf{Translation Adequacy:} how completely and correctly the spoken translation conveys the meaning of the source, judged by listening to the speech output.
  \item \textbf{Naturalness:} whether the delivery keeps a rhythm close to the source audio and to a real human simultaneous interpreter, i.e., whether pacing and rhythm track the source speaker and whether intonation is expressive rather than robotic or monotone.
  \item \textbf{Fidelity (audio quality):} whether the audio is clean and clear, free of noise, distortion, glitches, muffling, or artifacts.
\end{itemize}
For each sample, raters are given the source audio, its source transcript, and the generated target speech under test. Systems are anonymized and shuffled per sample (Figure~\ref{fig:human_eval_interface}). For a latency-matched comparison, we evaluate our \textbf{m5} tier and Seamless-Streaming's \textbf{thr\,=\,0.9} tier, whose LAAL is comparable to LiveInterpret~2.0's reported latency; we additionally report our lower-latency \textbf{m2} tier to characterize the quality--latency trade-off as latency tightens.

\begin{table*}[h]
  \centering
  \caption{Human MOS (1--5, higher is better) on RealSI sentence-level samples ($N{=}15$ per direction). $\pm$ denotes the sample standard deviation over the 15 rated items. Ours (m5) is latency-matched to LiveInterpret~2.0; Ours (m2) is a lower-latency additional operating point.}
  \label{tab:human_mos}
  \small
  \setlength{\tabcolsep}{4pt}
  \begin{tabular}{llcccc}
    \toprule
    \textbf{Direction} & \textbf{System} & \textbf{Naturalness} & \textbf{Fidelity} & \textbf{Adequacy} & \textbf{Overall} \\
    \midrule
    \multirow{4}{*}{En$\rightarrow$Zh}
      & Ours (m5) & 4.04$\pm$0.37 & 4.24$\pm$0.32 & 4.18$\pm$0.46 & 4.15$\pm$0.32 \\
      & Ours (m2) & 3.66$\pm$0.41 & 3.84$\pm$0.33 & 3.95$\pm$0.44 & 3.81$\pm$0.32 \\
      & LiveInterpret & 4.22$\pm$0.27 & 4.18$\pm$0.35 & 3.96$\pm$0.74 & 4.12$\pm$0.37 \\
      & Seamless-Streaming & 3.02$\pm$0.25 & 3.27$\pm$0.39 & 2.76$\pm$0.67 & 3.02$\pm$0.37 \\
    \midrule
    \multirow{4}{*}{Zh$\rightarrow$En}
      & Ours (m5) & 4.53$\pm$0.44 & 4.63$\pm$0.36 & 4.75$\pm$0.27 & 4.64$\pm$0.33 \\
      & Ours (m2) & 3.84$\pm$0.38 & 4.06$\pm$0.31 & 4.31$\pm$0.54 & 4.07$\pm$0.31 \\
      & LiveInterpret & 4.12$\pm$0.46 & 3.90$\pm$0.34 & 4.01$\pm$1.17 & 4.01$\pm$0.50 \\
      & Seamless-Streaming & 3.95$\pm$0.29 & 4.04$\pm$0.30 & 4.04$\pm$0.56 & 4.01$\pm$0.34 \\
    \bottomrule
  \end{tabular}
\end{table*}

\paragraph{Results.} As shown in Table~\ref{tab:human_mos}, at matched latency (m5), our system's Naturalness and Fidelity are top-tier in both directions, on par with or above LiveInterpret~2.0 (Overall 4.15 vs.\ 4.12 on En$\rightarrow$Zh, 4.64 vs.\ 4.01 on Zh$\rightarrow$En). At the lower-latency m2 tier scores drop, mainly due to cross-chunk discontinuities and a tighter delivery that feels less natural, though both tiers stay well above Seamless-Streaming. On Translation Adequacy, our m5 tier rates highest in both directions; since our raters are not professional interpreters, these scores may carry some bias, motivating the VIP LLM-as-judge protocol in Appendix~\ref{app:vip} as a complementary, less rater-dependent measure.

\paragraph{Statistical significance and agreement.} Table~\ref{tab:human_sig} reports paired Wilcoxon signed-rank tests comparing the Overall score of Ours (m5) against each external system per direction. Even at $N{=}15$, the rankings are not noise: on par with LiveInterpret~2.0 on En$\rightarrow$Zh (n.s.) and significantly better on Zh$\rightarrow$En ($p{=}0.003$), and significantly above Seamless-Streaming in both directions ($p{\le}0.001$, surviving Holm correction). Inter-annotator agreement is fair at the single-rater level (Krippendorff's $\alpha{=}0.43$ Overall, highest on Adequacy at $0.60$) and substantial for the aggregated 5-rater mean we actually compare (ICC(2,k)${=}0.76$ Overall; $0.77$/$0.72$ for En$\rightarrow$Zh/Zh$\rightarrow$En), indicating the ranking above reflects genuine quality differences rather than rater noise.

\begin{table}[h]
  \centering
  \caption{Paired Wilcoxon signed-rank test on the Overall human MOS, comparing Ours (m5) against each external system per direction.}
  \label{tab:human_sig}
  \small
  \setlength{\tabcolsep}{4pt}
  \resizebox{\columnwidth}{!}{%
  \begin{tabular}{llccc}
    \toprule
    \textbf{Direction} & \textbf{Comparison} & \textbf{Ours (m5)} & \textbf{Other} & \textbf{$p$} \\
    \midrule
    En$\rightarrow$Zh & vs.\ LiveInterpret & 4.15 & 4.12 & 0.93 (n.s.) \\
    En$\rightarrow$Zh & vs.\ Seamless-Streaming & 4.15 & 3.02 & 0.001 \\
    Zh$\rightarrow$En & vs.\ LiveInterpret & 4.64 & 4.01 & 0.003 \\
    Zh$\rightarrow$En & vs.\ Seamless-Streaming & 4.64 & 4.01 & 0.001 \\
    \bottomrule
  \end{tabular}%
  }
\end{table}

Due to time and resource constraints, we conduct only a small-scale human evaluation ($N{=}15$ per direction, $\sim$75 min per rater for the full 4-system session). Larger-scale evaluation with a broader and more diverse evaluator pool is left to future work.

\subsection{VIP: LLM-as-Judge Adequacy Evaluation}
\label{app:vip}

To assess translation quality at scale, we adopt \textbf{VIP}, an adequacy protocol originally proposed in the CLASI paper~\cite{cheng2024towards} for professional interpreters; we implement it with an LLM judge (DeepSeek-V4-Pro) over the ASR-transcribed output, rating Key Information Recognition, Correctness Assessment, and Expressiveness Assessment, and report the resulting pass rate (higher is better).

\begin{table*}[!h]
  \centering
  \caption{VIP pass rate (LLM-as-judge adequacy protocol; higher is better) across latency tiers $m1$--$m8$. LiveInterpret is the mean $\pm$ std over 3 official-API runs.}
  \label{tab:vip}
  \small
  \setlength{\tabcolsep}{4pt}
  \begin{tabular}{lccccccccc}
    \toprule
    \textbf{Direction} & \textbf{m1} & \textbf{m2} & \textbf{m3} & \textbf{m4} & \textbf{m5} & \textbf{m6} & \textbf{m7} & \textbf{m8} & \textbf{LiveInterpret} \\
    \midrule
    En$\rightarrow$Zh & 0.095 & 0.211 & 0.304 & 0.289 & 0.361 & 0.335 & 0.347 & 0.405 & 0.436$\pm$0.022 \\
    Zh$\rightarrow$En & 0.318 & 0.476 & 0.582 & 0.613 & 0.633 & 0.689 & 0.719 & 0.731 & 0.646$\pm$0.012 \\
    \bottomrule
  \end{tabular}
\end{table*}

As shown in Table~\ref{tab:vip}, VIP follows the same trend as our objective metrics and human MOS: quality rises steadily as latency loosens. On En$\rightarrow$Zh, a gap to LiveInterpret~2.0 remains (0.361 vs.\ 0.436 at m5), matching the ASR-BLEU trend in Table~\ref{tab:realsi_streaming_s2st_asr_bleu}. On Zh$\rightarrow$En, we reach parity from m5 (0.633 vs.\ 0.646) and surpass LiveInterpret~2.0 at higher tiers (up to 0.731 at m8).


\section{Case Study}
\label{app:case-study}

\begin{figure*}[htbp]
  \centering
  \includegraphics[width=0.98\linewidth]{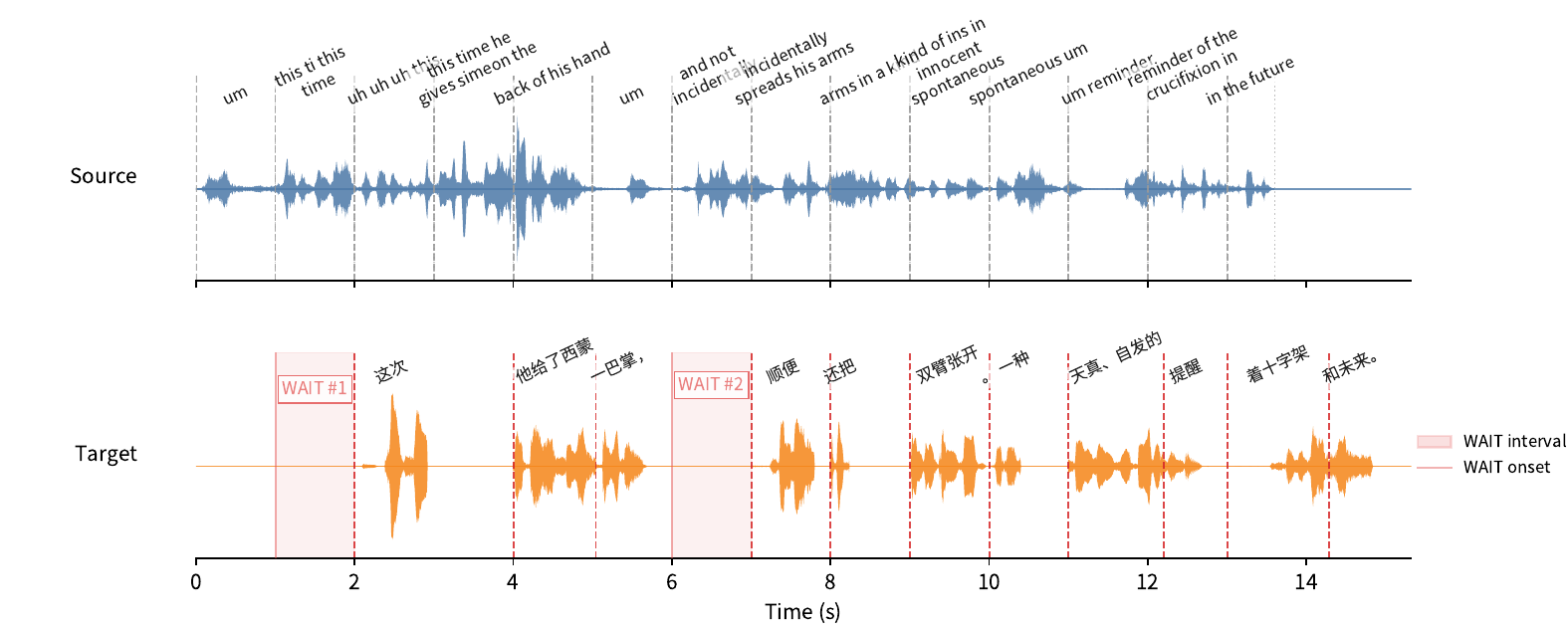}
  \caption{RealSI S2ST case study: En$\rightarrow$Zh, $m{=}1$.}
  \label{fig:case_en2zh_l1}
\end{figure*}

\begin{figure*}[htbp]
  \centering
  \includegraphics[width=0.98\linewidth]{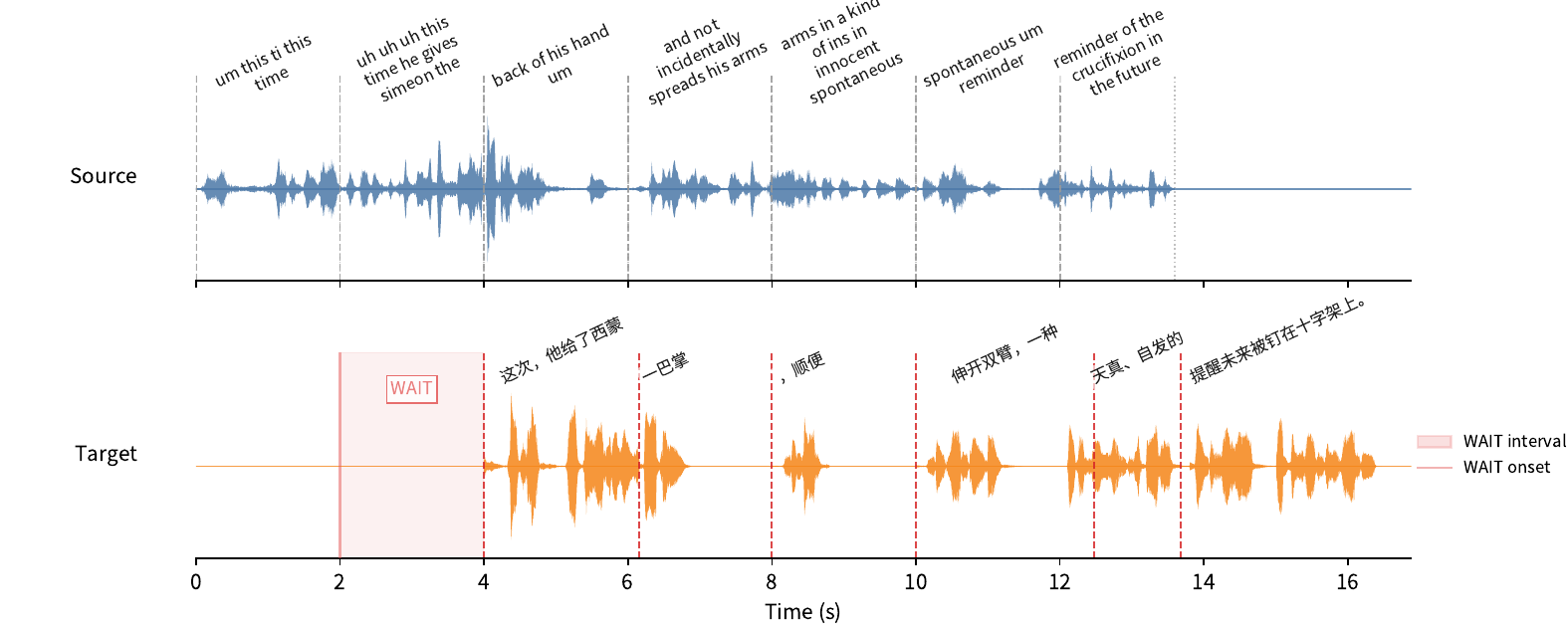}
  \caption{RealSI S2ST case study: En$\rightarrow$Zh, $m{=}2$.}
  \label{fig:case_en2zh_l2}
\end{figure*}

\begin{figure*}[htbp]
  \centering
  \includegraphics[width=0.98\linewidth]{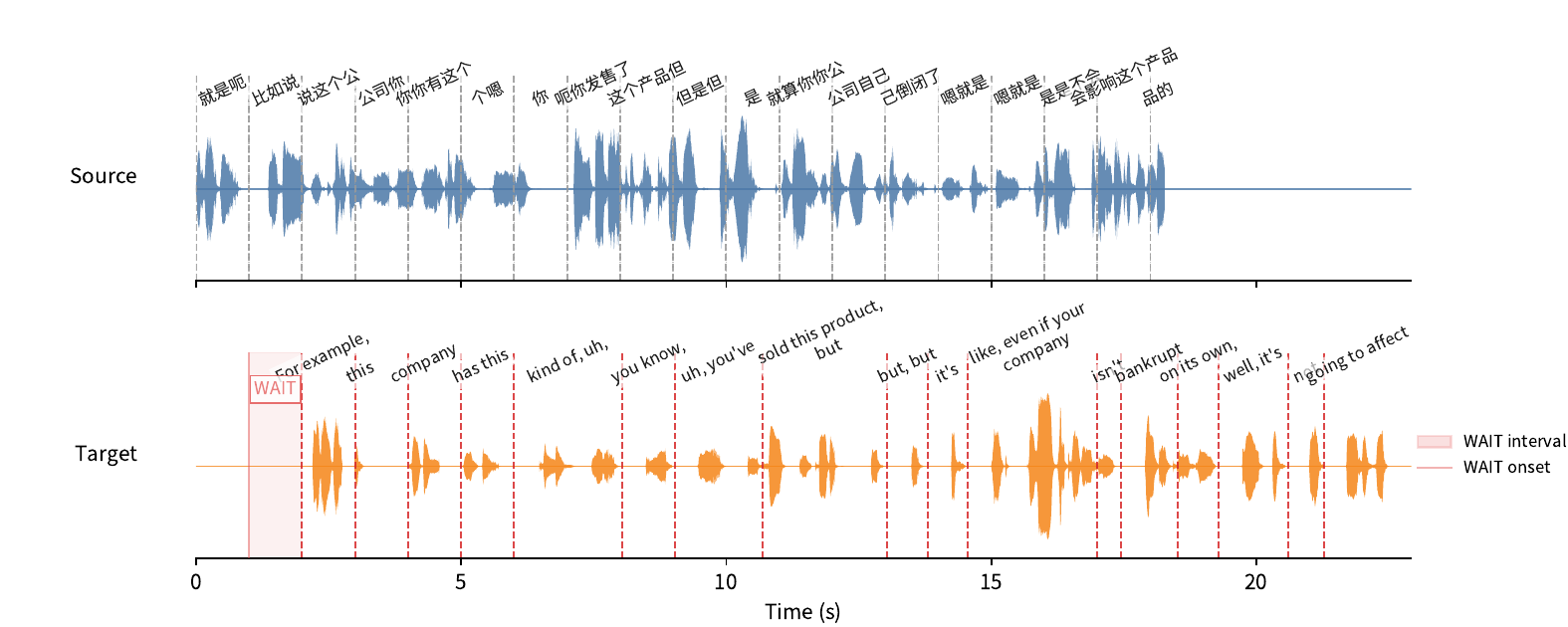}
  \caption{RealSI S2ST case study: Zh$\rightarrow$En, $m{=}1$.}
  \label{fig:case_zh2en_l1}
\end{figure*}

\begin{figure*}[htbp]
  \centering
  \includegraphics[width=0.98\linewidth]{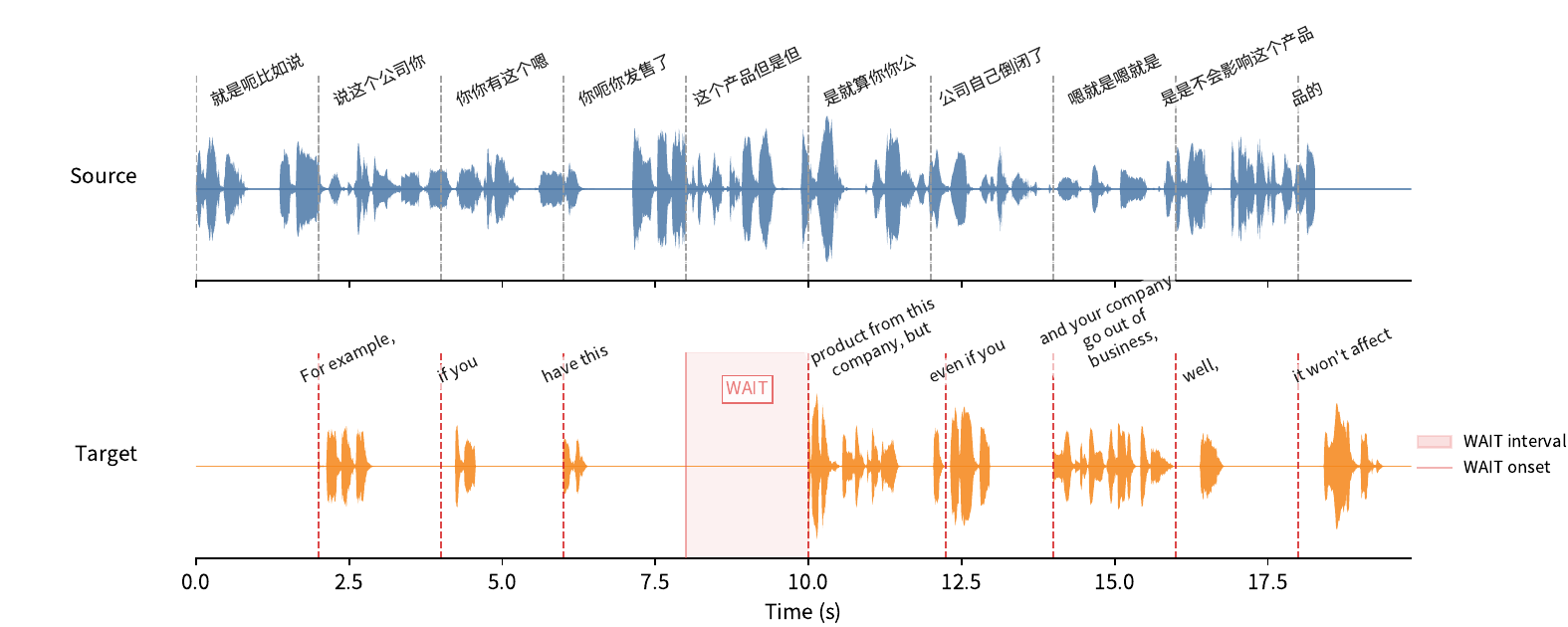}
  \caption{RealSI S2ST case study: Zh$\rightarrow$En, $m{=}2$.}
  \label{fig:case_zh2en_l2}
\end{figure*}

We complement the aggregate curves above with utterance-level visualizations of two RealSI sentence-level examples under two latency settings (Figures~\ref{fig:case_en2zh_l1}, \ref{fig:case_en2zh_l2}, \ref{fig:case_zh2en_l1}, and \ref{fig:case_zh2en_l2}).

Each figure aligns the source waveform (top) with the generated target waveform (bottom) on a shared time axis. Source words are aligned and grouped into $m$-second read windows (e.g., one second per step at $m{=}1$). Target segments denote partial translation commits. Pink \textit{WAIT} intervals indicate steps where the agent receives additional source audio but emits no speech, successfully accumulating context before the next write. 

At the finest-grained regime ($m{=}1$), the agent reads one source second per step, resulting in frequent but short target commits. As shown in the En$\rightarrow$Zh example, the system correctly predicts \textit{WAIT} steps early in the utterance, reflecting uncertainty at clausal boundaries during a disfluent source input (filled pauses and false starts). Raising the multiplier to $m{=}2$ changes the rhythm as expected: read windows widen, the number of target commits drops, and individual writes carry more context per step. This demonstrates the intended latency--quality trade-off of our framework.

\section{Bad Cases and Failure Analysis}
\label{app:bad-cases}

We identify three recurring failure modes encountered while constructing the paired streaming data and while decoding at tight latency. None of them affects the operating points reported in the main paper; rather, they characterize the current limits of our pipeline and motivate directions for future work.

\subsection{Boundary Concatenation Artifacts}
\label{app:bad-boundary}

\begin{figure*}[htbp]
  \centering
  \includegraphics[width=0.92\linewidth]{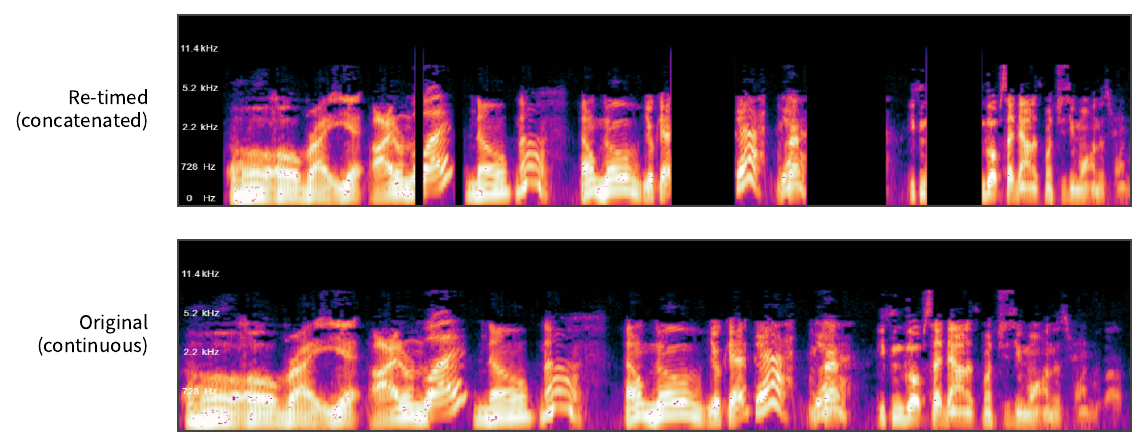}
  \caption{Boundary concatenation artifact (mel spectrogram). \emph{Top:} the causally re-timed target, showing hard cuts and silent gaps at chunk boundaries. \emph{Bottom:} the continuous original recording of the same content, with smooth onsets.}
  \label{fig:bad_boundary}
\end{figure*}

Figure~\ref{fig:bad_boundary} shows a representative \mbox{silence$\rightarrow$speech} join in a causally re-timed target. The artifact surfaces as a short transient click or spectral discontinuity at chunk boundaries, most audible at the utterance onset; it is primarily perceptual and only marginally raises ASR WER. It arises because voiced segments are hard-cut at systematically tight forced-alignment (FA) word boundaries and spliced against exact numerical silence, so the resulting join is an abrupt energy step rather than a natural onset.

\subsection{Forced-Alignment Error under Transcript--Audio Mismatch}
\label{app:bad-fa-mismatch}

\begin{figure*}[htbp]
  \centering
  \includegraphics[width=0.92\linewidth]{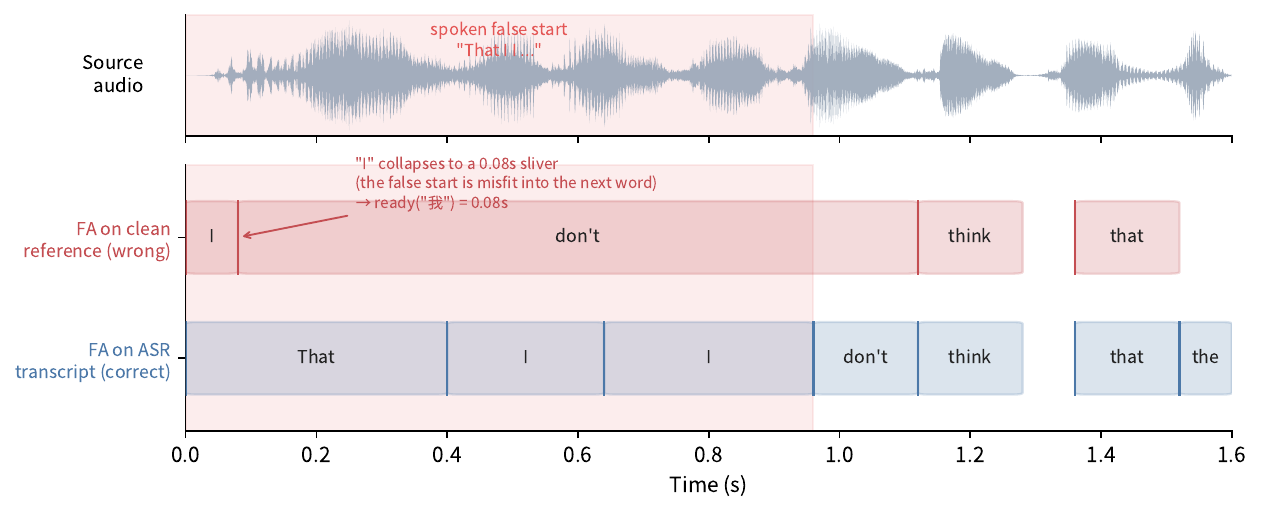}
  \caption{FA under transcript--audio mismatch (source clip; the false start \textit{``That I I''} is shaded). Aligning on the \textcolor[HTML]{C44E52}{clean reference} (top) collapses \textit{``I''} to an $0.08$\,s sliver, so the target \texttt{ready\_time} fires at $0.08$\,s; aligning on the \textcolor[HTML]{4C78A8}{ASR transcript} (bottom) matches the audio, where the real content only begins after $\sim\!0.96$\,s.}
  \label{fig:bad_fa_mismatch}
\end{figure*}

Figure~\ref{fig:bad_fa_mismatch} shows a clip whose clean reference transcript omits a spoken false start (\textit{``That I I\,\ldots''}), aligned once against the clean text and once against the ASR transcript. When aligned on the clean text, the aligner collapses the first source word \textit{``I''} into an $0.08$\,s sliver at the very onset and misattributes the false start to the following word; because this FA end-time becomes the cross-lingual \texttt{ready\_time}, the first target word is unlocked at $0.08$\,s and emitted almost immediately, before any real source content has arrived. This is a data issue rather than a modeling failure: when the transcript omits spoken material, FA must absorb it into neighboring words, and the scheduler then faithfully trusts these incorrect boundaries. We mitigate this by re-transcribing with ASR and filtering mismatched samples upstream.

\subsection{Premature Emission / Wrong Write}
\label{app:bad-premature}

\begin{figure*}[htbp]
  \centering
  \includegraphics[width=0.92\linewidth]{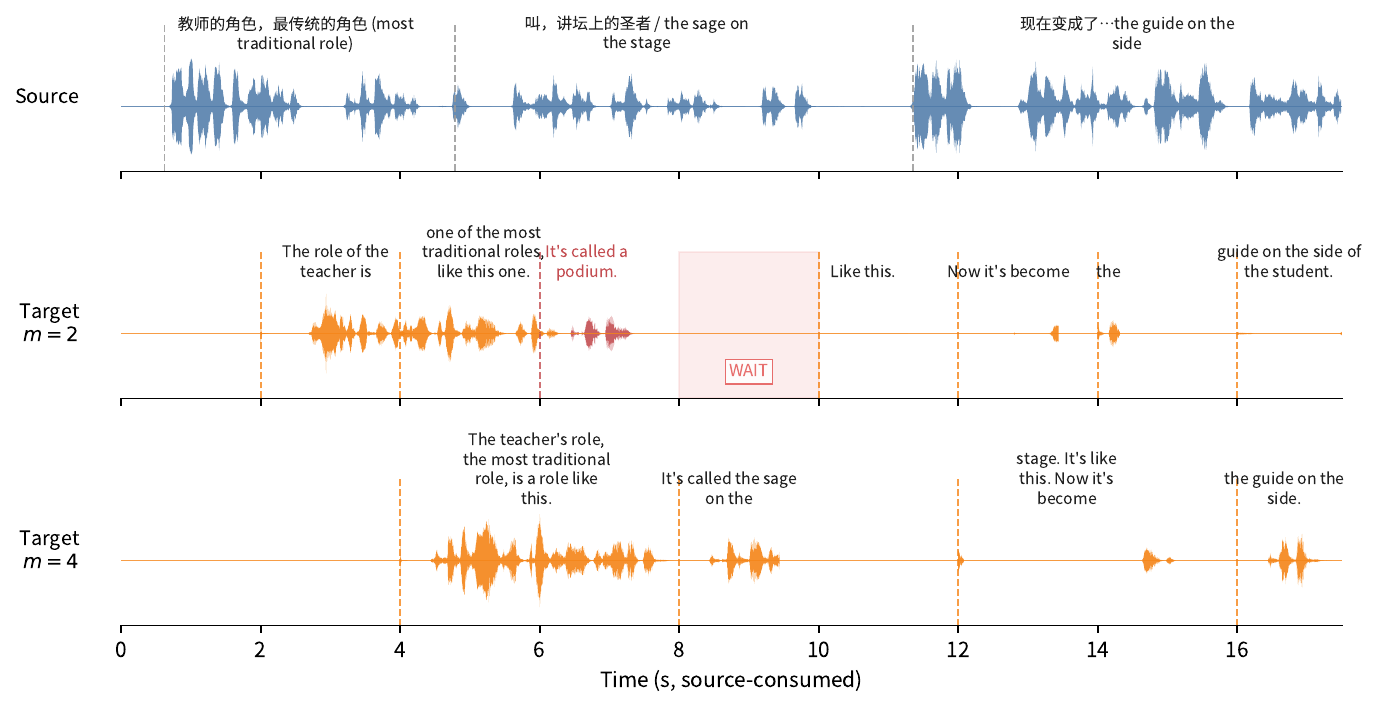}
  \caption{Premature emission / wrong write (Zh$\rightarrow$En), source-consumed axis. \emph{Top:} source. \emph{Middle ($m{=}2$):} the agent commits \textcolor[HTML]{C44E52}{\textit{``It's called a podium.''}} at $6.0$\,s---before the disambiguating source (``sage'') arrives---then stalls in a \textsc{wait}. \emph{Bottom ($m{=}4$):} the wider read window lets it hear the full phrase first and correctly emit \textit{``It's called the sage on the stage.''}}
  \label{fig:bad_premature}
\end{figure*}

Figure~\ref{fig:bad_premature} decodes the same Zh$\rightarrow$En utterance at a tight ($m{=}2$) and a looser ($m{=}4$) latency tier. At $m{=}2$, the agent commits \textit{``It's called a podium.''} at $6.0$\,s and then stalls in a \textsc{wait}; at $m{=}4$, it correctly emits \textit{``It's called the sage on the stage.''} At the tighter tier, the agent writes after hearing only the leading noun \textit{jiangtan} (``lectern''), before the disambiguating \textit{shengzhe} (``sage'') arrives, and cannot repair already-committed audio. This reflects the inherent read--write latency trade-off rather than a modeling defect.

\section{Reproducibility and Versioning}

All controlled comparisons share the same semantic-code tokenizer, acoustic backend, and evaluation scripts unless explicitly marked otherwise. The main manuscript reports the task mixture, paired-S2ST scale, decoding settings, and metric definitions; this appendix records the training stages, evaluation protocol, and representative configuration files needed to reconstruct the reported comparisons. Exact final release packaging, run identifiers, and artifact manifests will be listed in the public code release.

All public artifacts are used according to their original licenses and release terms.

An AI assistant was used during the paper-writing phase strictly for copy-editing, grammatical polishing, and LaTeX formatting of our drafted text. All core scientific ideas, experimental designs, and data analyses were conceptualized and executed entirely by the human authors.

\end{document}